\documentclass[12pt]{iopart}
\usepackage{graphicx}
\usepackage{graphics}

\begin{document}

\title{Sturmian bases for two-electron systems in hyperspherical coordinates}

\author{A. Abdouraman$^1$, Ana Laura Frappiccini$^2$, A. Hamido$^1$, F. Mota-Furtado$^3$, P. F. O'Mahony$^3$,  D. Mitnik$^4$, G. Gasaneo$^2$ and B. Piraux$^1$}

\address{
$^1$Institute of Condensed Matter and Nanosciences, Universit\'e Catholique de Louvain,\\ 
         2 chemin du cyclotron, Box L7.01.07,  B-1348 Louvain-la-Neuve, Belgium\\
$^2$Departamento de F\'isica, Universidad Nacional del Sur and CONICET, 8000 Bah\'ia Blanca, Buenos Aires, Argentina\\
$^3$Department of Mathematics, Royal Holloway, University of London, Egham, Surrey   TW20 0EX, United Kingdom\\
$^4$Instituto de Astronom\'ia y F\'isica del Espacio (IAFE,CONICET-UBA), Argentina }

\begin{abstract}
We give a detailed account of an {\it ab initio} spectral approach for the calculation of  energy spectra of two active electron atoms in a system of hyperspherical coordinates.
In this system of coordinates, the Hamiltonian has the same structure as the one of atomic hydrogen with the Coulomb potential expressed in terms of a hyperradius and
the nuclear charge replaced by an angle dependent effective charge. The simplest spectral approach consists in expanding the hyperangular  wave function in a basis of hyperspherical harmonics. This expansion however, is known to be very slowly converging. Instead, we introduce new hyperangular sturmian functions. These functions do not have an analytical expression but they treat the first term of the multipole expansion of the electron-electron interaction potential, namely the radial electron correlation, exactly. The properties of these new functions are discussed in detail. For the basis functions of the hyperradius, several choices are possible. In the present case, we use Coulomb sturmian functions of half integer angular momentum. We show that, in the case of H$^-$, the accuracy of the energy and the width of the resonance states  obtained  through a single diagonalization of the Hamiltonian, is comparable to the values given by state-of-the-art methods while using a much smaller basis set. In addition, we show that precise values of the electric-dipole oscillator strengths for $S\rightarrow P$ transitions in helium are obtained thereby confirming the accuracy of the bound state wave functions generated with the present method.  
\end{abstract}
\submitto{J. Phys. B: At. Mol. Opt. Phys.}
\maketitle

\section{INTRODUCTION}

The use of hyperspherical coordinates in describing strongly correlated two-electron atomic systems has provided a deep insight into electron dynamics. In this system of coordinates, the structure of the Hamiltonian is the same as for atomic hydrogen:  the electrostatic potential can be written as $C(\Omega)/R$ where $R$ is an hyperradius which determines the size of the system and $C(\Omega)$ is an effective charge depending on a hyperangle describing radial correlations, and on the polar and azimuthal angles of both electrons. Initially, Bartlett \cite{Bartlett} and Fock \cite{Fock} were the first to study with this system of coordinates, the ground state wave function of helium and to derive a series expansion, known as the "Fock expansion" that accounts for all the singularities occurring when both electrons are close to each other and when each electron is close to the nucleus. Later on, and in order to study  doubly excited states of He and H$^-$, Macek \cite{Macek1} introduced an adiabatic expansion that has the same form as a Born-Oppenheimer expansion providing a convenient framework to elucidate the symmetry of these doubly excited states (see also \cite{Fano}). Although most of the theoretical work based on this approach \cite{Macek1,Lin1,Lin2} has mainly focused on qualitative interpretation of two-electron processes, it has also produced quantitative results on the energy eigenstates of He and H$^-$ as well as on cross-sections for photon and electron impact collisions with atoms \cite{Starace1}. Very accurate results have been obtained by including non-adiabatic coupling terms, making the efficiency of this method at least comparable to other highly sophisticated approaches \cite{Lin3}.\\

Spectral approaches employing hyperspherical coordinates have also been used to solve directly both the stationary and the time-dependent Schr\"odinger equation (see for instance \cite{Watanabe}). The simplest method is based on expanding the hyperangular  wave function in a basis of hyperspherical harmonics \cite{Das}. Such spectral approaches are particularly indicated in the treatment of electron impact ionization of atomic hydrogen and double photoionization of two-electron atomic systems since in the hyperspherical system of coordinates, the double continuum wave function takes a very simple analytical form when the electrons are both asymptotically far from the residual ion \cite{Rudge}. However, it turns out that the convergence of such an expansion in hyperspherical  harmonics is usually very slow \cite{Knirk,Klar}. In fact, hyperspherical harmonics are not suitable for describing a situation where one electron is very far from the other.\\

In order to study low-energy fragmentation states of three-charged particle systems, and in particular, Wannier's threshold laws, Macek and Ovchinnikov  \cite{Macek2} replaced the basis of hyperspherical harmonics by a very small size basis of what they called angle sturmians. These functions that depend on all the angles, are orthonormal with respect to the effective charge $C(\Omega)$ and, as a result, describe very well the angular motion of the electrons in situations where one electron stays confined while the second one moves slowly away. In the present contribution, we follow a similar idea and introduce a basis of hyperangular sturmian functions of the hyperangle. These functions are solutions of a Sturm-Liouville equation in which the weight function is an effective charge that includes the electron-nucleus interaction potentials and the first term of the multipole expansion of the electron-electron interaction term.  Those functions form a complete and discrete set of $\mathcal{L}^2$-integrable functions that are orthogonal with respect to this effective charge. By comparison, hyperspherical harmonics whose dependence on the hyperangle is expressed in terms of Jacobi polynomials, are solutions of a Sturm-Liouville equation that does not involve any effective charge. As a matter of fact, these functions give a poor description of the electron angular motion.\\

In this contribution, we show that  accurate values of the bound state eigenenergies and of  the energy and width of the doubly excited states for $He$ and $H^-$ may be obtained from a single diagonalization of the atomic hamiltonian calculated in a sturmian basis of moderate size of which is significantly smaller than for example a basis set of hyperspherical harmonics. In addition, in order to assess the accuracy of the bound state wave functions, we calculate the electric-dipole oscillator strengths for $S\rightarrow P$ transitions in helium. Our results obtained in the length gauge demonstrate that the wave functions are very accurate even at large distances.\\

 The paper is divided in two sections. The first one is devoted to the general theory. After briefly reviewing the hyperspherical coordinates, we introduce our sets of  hyperangular and hyperradial sturmian functions used for the spectral analysis of the solution of the time independent Schr\"odinger equation with or without complex rotation of the Hamiltonian. The reasons why hyperangular sturmians are more appropriate than Jacobi polynomials, are discussed in great detail. We also compare sturmian expansions in spherical and hyperspherical coordinates. In particular, we show that the optimization of the sturmian basis in terms of free nonlinear parameters may be carried out in both systems of coordinates. Results for the energy of bound states, energy and width of doubly excited states for $He$ and $H^-$ and electric-dipole oscillator strengths in the length gauge for $S\rightarrow P$ transitions in $He$ are presented, discussed and compared to very accurate existing data. To conclude, we show in what context the use of such a basis of hyperangular sturmians is pertinent. Atomic units are used throughout
unless otherwise specified.\\

\section{THEORY}
\subsection{Basic formulae}
The wave function of a two-electron atomic system with total angular momentum $L$, $L_z$ component $M$, and total energy $E_{\beta}$, satisfies  
the following stationary Schr\"odinger equation:
\begin{equation}
(H_I +U)\Phi^{L,M}_{\beta}(\vec{r_1},\vec{r_2})=E_{\beta}\Phi^{L,M}_{\beta}(\vec{r_1},\vec{r_2}).
\label{eq1}
\end{equation}
where $\vec{r}_1$ and $\vec{r}_2$ are the position vectors of both electrons with respect to the nucleus. $H_I$ is the independent electron 
Hamiltonian which is:
\begin{equation}
H_I= -\frac{1}{2}\triangle_{r_1}-\frac{1}{2}\triangle_{r_2}-Z\left( \frac{1}{r_1}+\frac{1}{r_2}\right) , 
\end{equation}
where Z denotes the charge of the nucleus which is assumed to be infinitely massive. $U$ is the electron-electron interaction Hamiltonian
which can be expressed in terms of its well known multipole expansion as follows:
\begin{equation}
U=\frac{1}{|\vec{r}_1-\vec{r}_2|}=\sum_{q=0}^{\infty}\sum_{p=-q}^q\frac{4\pi}{2q+1}\frac{r^{q}_{<}}{r^{q+1}_{>}}Y^*_{q,p}(\theta_1,\phi_1)Y_{q,p}(\theta_2,\phi_2),
\end{equation}
where $r_<$=min($r_1,r_2$) and $r_>$=max($r_1,r_2$) with $(r_i,\theta_i,\phi_i)$ the spherical coordinates of the position vector $\vec{r}_i$ ( $i=1,2$). 
Let us consider the hyperspherical coordinate system which are related to the spherical coordinates as follows:
\begin{equation}
(r_1,r_2,\theta_1,\theta_2,\phi_1,\phi_2)\rightarrow(R,\alpha,\theta_1,\theta_2,\phi_1,\phi_2),
\end{equation}
where $r_1$ and $r_2$ are replaced by the hyper-radius $R$ and the hyper-angle $\alpha$
 \begin{eqnarray}
 R&=&\sqrt{r_1^2+r_2^2},\\
  \alpha&=&\arctan\left( \frac{r_2}{r_1}\right) .
 \end{eqnarray}
In hyperspherical coordinates, the kinematics of three particles is reduced to that of the motion of one body of mass
$\mu$  on the five-dimensional surface of a six-dimensional sphere whose variable radius is R. R varies from zero to infinity and the hyperangular coordinate $\alpha$  
varies between 0 and $\pi/2$. The reduced mass $\mu$ is defined as $\mu=\sqrt{m_i m_j m_k /(m_i +m_j +m_k)}$ where $m_i$, $m_j$ and $m_k$ are the
masses of the three particles.  For the case under consideration here - one nucleus and two electrons - the reduced mass is equal to 1 in the limit of an infinitely 
massive nucleus.\\

In this system of coordinates, the wave function of the two-electron atomic system  satisfies the stationary  Schr\"odinger equation: 
 \begin{equation}
(H_I +U)\Phi^{L,M}_{\beta}(R,\alpha,\mbox{\^{r}}_1,\mbox{\^{r}}_2)=E_{\beta}\Phi^{L,M}_{\beta}(R,\alpha,\mbox{\^{r}}_1,\mbox{\^{r}}_2),    \label{SE}
\end{equation}
 where the independent electron Hamiltonian $H_I$ and the electron-electron interaction become respectively:
\begin{eqnarray}
H_I&=& -\frac{1}{2}\frac{\partial^2}{\partial R^2}-\frac{5}{2R}\frac{\partial}{\partial R}+\frac{\Lambda^2}{2R^2}-\frac{Z}{R}\left( \frac{1}{\cos{\alpha}}+\frac{1}{\sin{\alpha}}\right) , \\ 
\nonumber\\  
U&=&\frac{1}{R}\sum_{q=0}^{\infty}\sum_{p=-q}^q\frac{4\pi}{2q+1}v_q(\alpha)Y^*_{q,p}(\theta_1,\phi_1)Y_{q,p}(\theta_2,\phi_2). 
\end{eqnarray}
The Casimir operator $\Lambda^2$ appearing in Eq. (8) is given by:
\begin{equation}
\Lambda^2=-\frac{1}{\sin^2\alpha\;\cos^2\alpha}\;\frac{\partial}{\partial\alpha}\left(\sin^2\alpha\;\cos^2\alpha\;\frac{\partial}{\partial\alpha}\right)+\frac{\hat l_1^2}{\cos^2\alpha}+\frac{\hat l_2^2}{\sin^2\alpha},   \label{Casimir}
\end{equation}
where $\hat l_1^2$ and $\hat l_2^2$ are the individual electron angular momentum operators. In Eq. (9), the factors $v_q(\alpha)$ are:
\begin{equation}
 v_q(\alpha)= \left\{ \begin{array} {cc} 
\frac{(\tan\alpha)^q}{\cos{\alpha}}  &\mbox{if $ 0<\alpha<\frac{\pi}{4}$};   \\ \\ 
\frac{(\cot\alpha)^q}{\sin{\alpha} }  &\mbox{if $\frac{\pi}{4}<\alpha<\frac{\pi}{2}$}.
\end{array} \right.
\end{equation}
In this system of coordinates, the electron-electron and the electron-nucleus Coulomb interactions are reduced to a hyperspherical Coulomb interaction $C(\alpha,\hat{r}_1,\hat{r}_2)/R$ where $C(\alpha,\hat{r}_1,\hat{r}_2)$ can be considered as an effective charge depending on the angles only:
\begin{eqnarray}
C(\alpha,\hat{r}_1,\hat{r}_2)=&-&Z\left(\frac{1}{\cos{\alpha}}+\frac{1}{\sin{\alpha}}\right)\nonumber\\
&+&\sum_{q=0}^{\infty}\sum_{p=-q}^q\frac{4\pi}{2q+1}v_q(\alpha)Y^*_{q,p}(\theta_1,\phi_1)Y_{q,p}(\theta_2,\phi_2).
\end{eqnarray}

In order to eliminate the first derivatives both in $R$ and $\alpha$ in the Hamiltonian (8) and the Casimir operator (\ref{Casimir}), it is convenient to introduce the wave function transformation:  
\begin{equation}
\Phi^{L,M}_{\beta}(R,\alpha,\mbox{\^{r}}_1,\mbox{\^{r}}_2)=\frac{1}{R^{\frac{5}{2}}\sin{\alpha}\cos{\alpha}}\;\Psi^{L,M}_{\beta}(R,\alpha,\mbox{\^{r}}_1,\mbox{\^{r}}_2).  \label{Psi}
\end{equation}
With this change of function the Schr\"odinger equation (\ref{SE}) is transformed into the following equation for $\Psi^{L,M}_{\beta}(R,\alpha,\mbox{\^{r}}_1,\mbox{\^{r}}_2)$:
\begin{equation}
\left[  -\frac{1}{2}\frac{\partial^2}{\partial R^2}+\frac{\tilde\Lambda^2+2 R C(\alpha,\hat{r}_1,\hat{r}_2)}{2R^2}-E_{\beta}\right]\Psi^{L,M}_{\beta}(R,\alpha,\mbox{\^{r}}_1,\mbox{\^{r}}_2) =0, \label{Reduced}
\end{equation}
where the reduced Casimir operator $\tilde\Lambda^2$ is:
\begin{equation}
\tilde\Lambda^2=-\frac{\partial^2}{\partial\alpha^2}-\frac{1}{4}+\frac{\hat l_1^2}{\cos^2\alpha}+\frac{\hat l_2^2}{\sin^2\alpha}. \label{ReduCasimir}
\end{equation}

\subsection{Spectral analysis of the solution}
In order to solve Eq. (\ref{Reduced}), for $\Psi^{L,M}_{\beta}(R,\alpha,\mbox{\^{r}}_1,\mbox{\^{r}}_2)$ given in Eq. (\ref{Psi}) we use the following expansion:
\begin{eqnarray}
\Psi^{L,M}_{\beta}(R,\alpha,\mbox{\^{r}}_1,\mbox{\^{r}}_2)&=&\sum_{n,p,l_1,l_2}\frac{1}{\sqrt{2(1+\delta_{l_1,l_2})}}\psi^{L,M}_{n,p,l_1,l_2} S_{n,p}(R)\nonumber\\
&\times&\left(\phi_{p,l_1,l_2}^{L,M}(\alpha,\hat r_1,\hat r_2)+
(-1)^S\phi_{p,l_1,l_2}^{L,M}(\frac{\pi}{2}-\alpha,\hat r_2,\hat r_1)\right) \label{Expansion}
\end{eqnarray}
where we take into account the exchange of the electrons. $S$ is the total electron spin taking the value 0 and 1 for singlet and triplet states respectively. To solve Eq. (\ref{Reduced}) any type of basis function can be used both for $S_{n,p}(R)$ and $\phi_{p,l_1,l_2}^{L,M}(\alpha,\hat r_1,\hat r_2)$. However, we choose here to use sturmian functions both for the hyperradial and the hyperangular parts.

\subsubsection{Hyperangular sturmian functions}

An apparently natural definition for the hyperangular basis functions results from the definition  of the reduced Casimir operator given by Eq. (\ref{ReduCasimir}). In many mathematical 
books \cite{Morse and Feshback 1953}, it is found that the eigenfunctions $\phi_{p,l_1,l_2}^{L,M}(\alpha,\hat r_1,\hat r_2)$ of  $\tilde\Lambda^2$ satisfy the following eigenvalue problem:
\begin{equation}
\tilde\Lambda^2\phi_{p,l_1,l_2}^{L,M}(\alpha,\hat r_1,\hat r_2)=
\left( \lambda(\lambda+4)+\frac{15}{4}\right)
 \phi_{p,l_1,l_2}^{L,M}(\alpha,\hat r_1,\hat r_2),  \label{AngularJEquation}
\end{equation}
and  are written in terms of bipolar harmonics $Y^{L,M}_{l_1,l_2}(\hat r_1,\hat r_2)$ and Jacobi polynomials $P^{l_1,l_2}_{p}(\cos 2\alpha)$ :
\begin{equation}
\phi_{p,l_1,l_2}^{L,M}(\alpha,\hat r_1,\hat r_2) =(\cos\alpha)^{l_1+1}(\sin\alpha)^{l_2+1}P^{l_1,l_2}_{p}(\cos 2\alpha)Y^{L,M}_{l_1,l_2}(\hat r_1,\hat r_2).   \label{JacobiBipolar}
\end{equation}
where $\lambda=2 p+l_1+l_2$ is a non-negative integer.
These functions which are the so-called hyperspherical harmonics, are the eigenfunctions of the total angular momentum of the system, but they do not include any of the correlation produced by the Coulomb interaction. For that reason, the convergence of expansion (\ref{Expansion}) is very slow. In order to overcome this problem, we define here an hyperangular sturmian basis that includes most of the correlation produced 
by the three-body Coulomb interactions.\\

Similarly to what was done by Macek and Ovchinnikov \cite{Macek2} and \cite{Gasaneo2009,Gasaneo2012} and instead of solving the simple  Eq. (\ref{AngularJEquation}) we consider the following  
equation: 
\begin{equation}
\left[  \tilde\Lambda^2+2  \rho \, \tilde C(\alpha,\hat{r}_1,\hat{r}_2)\right]\phi_{p,l_1,l_2}^{L,M}(\alpha,\hat r_1,\hat r_2) 
=\left( \lambda(\lambda+4)+\frac{15}{4}\right)
\phi_{p,l_1,l_2}^{L,M}(\alpha,\hat r_1,\hat r_2), \label{AngSturmians1}
\end{equation}
which results from the Schr\"odinger Eq. (\ref{Reduced}) by replacing $R$ by $\rho$, $Z$ by $\mathcal{Z}$ and  $C(\alpha)$ by $\tilde{C}(\alpha)$ which we will call the reduced effective charge and which is defined as follows:
\begin{equation}
\tilde{C}(\alpha)=-\mathcal{Z}\left(\frac{1}{\cos{\alpha}}+\frac{1}{\sin{\alpha}}\right)
+v_0(\alpha). \label{CalphaTilde}
\end{equation}
We write the eigensolution $\phi_{p,l_1,l_2}^{L,M}(\alpha,\hat r_1,\hat r_2)$ as:
\begin{equation}
\phi_{p,l_1,l_2}^{L,M}(\alpha,\hat r_1,\hat r_2) =H^{l_1,l_2}_{p,\mathcal{Z},\lambda}(\alpha)Y^{L,M}_{l_1,l_2}(\hat r_1,\hat r_2),   \label{SturBipolar}
\end{equation}
where $H^{l_1,l_2}_{p,\mathcal{Z},\lambda}(\alpha)$ satisfies the following Sturm-Liouville equation:
\begin{eqnarray}
[-\frac{\partial^2}{\partial\alpha^2}-\frac{1}{4}+\frac{l_1(l_1+1)}{\cos^2\alpha}+\frac{l_2(l_2+1)}{\sin^2\alpha}&-&(\lambda(\lambda+4)+\frac{15}{4})\;\;]H^{l_1,l_2}_{p,\mathcal{Z},\lambda}(\alpha)=\nonumber\\
&-&2\rho_p(\lambda) \tilde{C}(\alpha)H^{l_1,l_2}_{p,\mathcal{Z},\lambda}(\alpha), \label{AngSturmian}
\end{eqnarray}
with the boundary conditions:
\begin{eqnarray}
\lim_{\alpha\rightarrow 0} H_{p,\mathcal{Z},\lambda}^{l_1,l_2}(\alpha)&=&0,\\
\lim_{\alpha\rightarrow \frac{\pi}{2}} H_{p,\mathcal{Z},\lambda}^{l_1,l_2}(\alpha)&=&0. \label{BounConditions}
\end{eqnarray}
In Eq. (\ref{AngSturmian}), $\lambda$ is  considered as a fixed parameter, $2 \rho_p(\lambda)$ is the eigenvalue and $\tilde{C}(\alpha)$ is the weight function.  The functions $H^{l_1,l_2}_{p,\mathcal{Z},\lambda}(\alpha)$ form an orthogonal and complete set of functions:
\begin{equation}
\int_0^{\frac{\pi}{2}}H^{l_1,l_2}_{q,\mathcal{Z},\lambda}(\alpha)\, \tilde{C}(\alpha) \,H^{l_1,l_2}_{p,\mathcal{Z},\lambda}(\alpha) \;\mathrm{d}\alpha=\delta_{p,q},  \label{Horto}
\end{equation}
and
\begin{equation}
\sum_p H^{l_1,l_2}_{p,\mathcal{Z},\lambda}(\alpha')\, \tilde{C}(\alpha) \,H^{l_1,l_2}_{p,\mathcal{Z},\lambda}(\alpha) =\delta(\alpha-\alpha').  \label{HClosure}
\end{equation}
To within a scale factor ($1/R$), the reduced effective charge includes the electron-nucleus interaction and the first term $v_0(\alpha)$ of the multipole expansion of the electron-electron interaction U 
given by Eq. (9). $v_0(\alpha)$ is in fact the radial electronic repulsion. Because $\tilde{C}(\alpha)$ is the weight function in the orthonormality relation (\ref{Horto}), it must have a fixed sign in its domain of definition. This implies that none of the higher order terms of the multipole expansion of U can be included in $\tilde{C}(\alpha)$  without affecting the  positive definite character of the weight for all values of  $\alpha$.  This contrasts with the approach of Macek and Ovchinnikov where the full effective charge $C(\alpha,\hat{r}_1,\hat{r}_2)$ of Eq. (\ref{Reduced}) was considered.
It is important to note that our hyperangular sturmian functions  have no known analytical solution. Instead, it is necessary to derive them numerically. In addition to the fully numerical approach developed in \cite{Gasaneo2009,Gasaneo2012}, we use in this contribution two spectral methods based on an expansion in Tchebychev polynomials and also an expansion in terms of B-splines. It turned out that the B-spline expansion is the most accurate of the spectral methods considered here. This expansion and the problem of the exchange of the electrons are treated in the next section.\\

Let us now discuss in more detail why the hyperangular sturmian basis is by far, more appropriate than the Jacobi polynomial basis in the present case.  There are essentially two reasons:  the hyperangular 
sturmians take explicitly into account both the electron-nucleus and the electron-electron interactions. In general, in an atomic two-electron system, at least one of the electrons is localized around $r_1\simeq0$ or $r_2\simeq0$ in the bound states, the single continua  and the excitation-ionization channels. In these cases and for most values of $R$, the corresponding states  will be localized close to $ \alpha\simeq0$ or $\alpha\simeq\pi/2$. This means that somehow we need to concentrate the nodes of the hyperangular eigenfunctions in those regions of  $\alpha$. The Jacobi polynomials do not fulfill this condition; their nodes are spread almost uniformly along the  $\alpha$ axis between 0 and $\pi/2$. Taking into account $\tilde{C}(\alpha)$ given by Eq. (\ref{CalphaTilde}) in Eq. (22) moves the nodes to the inner regions close to $0$ or $\pi/2$. \\

\begin{figure}[h]
\includegraphics[width=12cm,height=9cm]{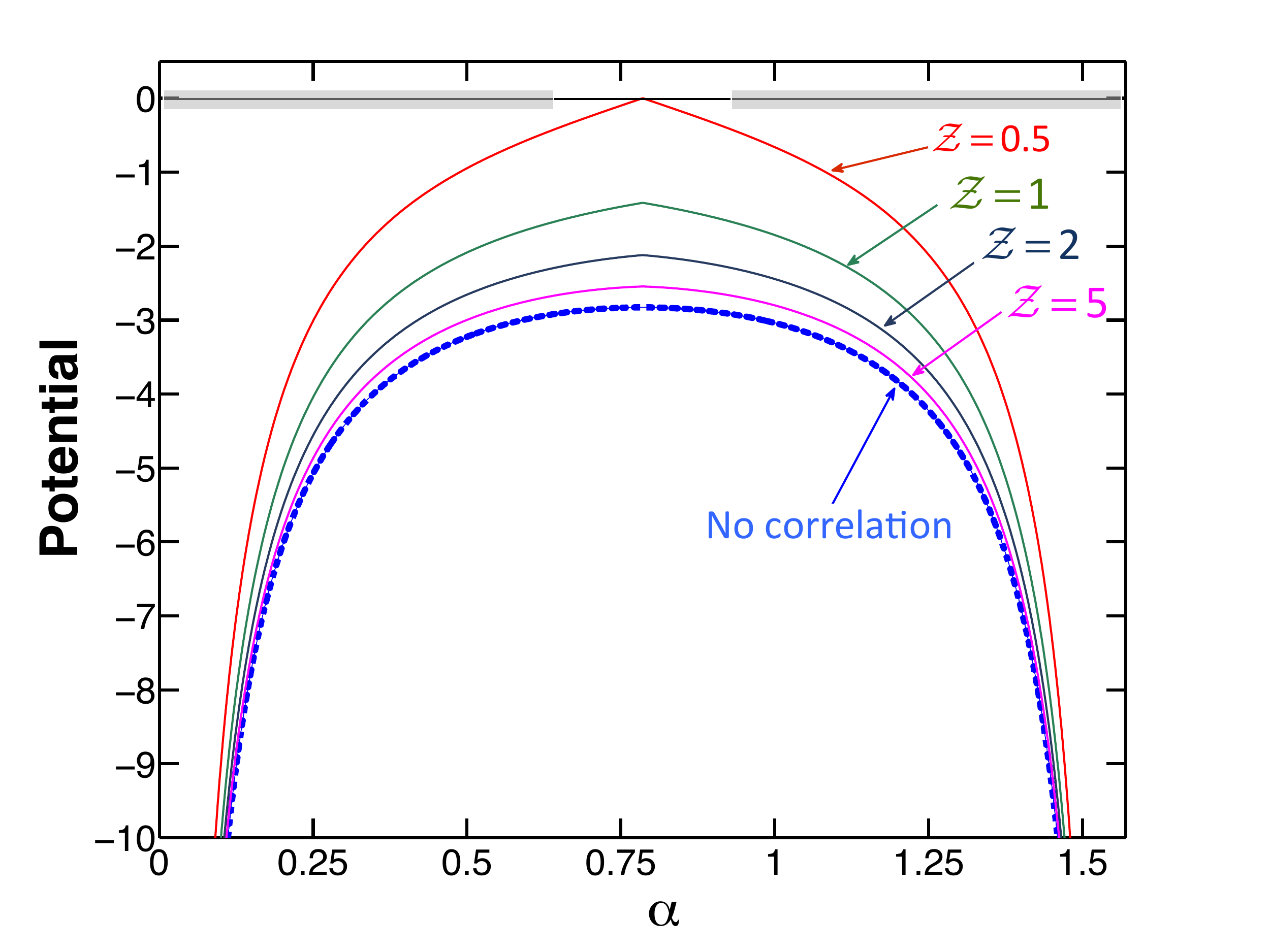}
\caption{ (Color on line) Plot of the potential term in the Schr\"odinger equation (27). This potential term is actually  the effective charge $\tilde{C}(\alpha)$ represented here for various values of 
	         $\mathcal{Z}$. The dashed (blue) line is obtained by neglecting $v_0(\alpha)$, the radial electronic repulsion. The nodes of the hyperangular  sturmians concentrate in the shaded area on the 
	         horizontal axis.}
\label{Fig1}
\end{figure}
It is quite easy to understand why the inclusion of the electron-nucleus and the electron-electron interactions affects the location of the nodes of the hyperangular sturmian functions. To make this clear, let us consider Eq. (\ref{AngSturmian}) in the case where $l_1=l_2=0$:
\begin{equation}
\left[-\frac{\partial^2}{\partial\alpha^2}- \epsilon_a\right]H^{0,0}_{p,\mathcal{Z},\lambda}(\alpha)=-2\rho_p
\left[-\mathcal{Z}\left(\frac{1}{\cos{\alpha}}+\frac{1}{\sin{\alpha}}\right)
+v_0(\alpha) \right] H^{0,0}_{p,\mathcal{Z},\lambda}(\alpha), \label{AngSturmianl0}
\end{equation}
where we defined $\epsilon_a=\frac{1}{4}+\left(\lambda(\lambda+4)+\frac{15}{4}\right)$. The boundary conditions (23) and (24) force $H^{0,0}_{p,\mathcal{Z},\lambda}(\alpha)$ to be zero at the points $0$ and $\pi/2$ allowing us to associate Eq. (\ref{AngSturmianl0}) with the well known quantum problem of a particle in a well with infinite walls. Let us suppose for the time being that we set $\rho_p=0$ in (\ref{AngSturmianl0}). In that case, equation (27) becomes:
\begin{equation}
\left[-\frac{\partial^2}{\partial\alpha^2}- \epsilon_a\right]H^{0,0}_{p,\mathcal{Z},\lambda}(\alpha)=0, 
\end{equation}
 which is the same as the Schr\"odinger  equation describing a free particle inside an infinite well where $\epsilon_a$ is the energy eigenvalue. It is clear that the first function ($p=0$) has no node, the second does have one, and the number of nodes grows as the eigenvalue grows. The nodes are regularly distributed along the region of $\alpha$ as if it were a string with both ends fixed. Now, let us take, e.g. $\rho_p=1$. Then, instead of having a particle in a simple well with infinite wall, we have still the walls but inside the well the potential varies as shown in Fig. (\ref{Fig1}). In the figure we plotted the potential  for different values of $\mathcal{Z}=0.5,1,2$ and $5$. In the plot we included also the potential corresponding to the independent electron (no correlation) model. As the potential becomes more attractive near the borders, the kinetic energy increases significantly leading to fast oscillations of the wave function. That means that the nodes are going to be mainly localized closer to the borders of the wells in the shaded area in Fig. (1). Besides, when comparing the potentials in Fig. (\ref{Fig1}), we notice that the width of the potential wells decreases with increasing $\mathcal{Z}$ thereby implying a bigger concentration of nodes close to $\alpha=0$ and $\alpha=\pi/2$. The case where $\mathcal{Z}\simeq 0.5$ is when the potential is 0 at $\pi/4$. This is the extreme situation where the nodes are pushed away from the center and concentrate at the edges of the angular region. \\  
 
In the previous description we assumed $\epsilon_a$ as being the eigenvalue. However, in the methodology we are implementing, we are taking $\rho_p$ as the eigenvalue keeping $\epsilon_a=\frac{1}{4}+\left(\lambda(\lambda+4)+\frac{15}{4}\right)$ constant. This is what turns the $H^{l_1,l_2}_{p,\mathcal{Z},\lambda}(\alpha)$ from energy eigenfunctions to angular sturmian functions. When keeping fixed $\frac{1}{4}+\left(\lambda(\lambda+4)+\frac{15}{4}\right)$ we are looking for all the potential strengths possessing that energy. This is similar to what we do with the Coulomb potential. If we fix the charge, the Schr\"odinger equation provides the energies. If we fix the energy, the equation gives all the charges whose corresponding potential is able to support one state of the given energy.\\

\begin{figure}[h]
\includegraphics[width=12cm,height=9cm]{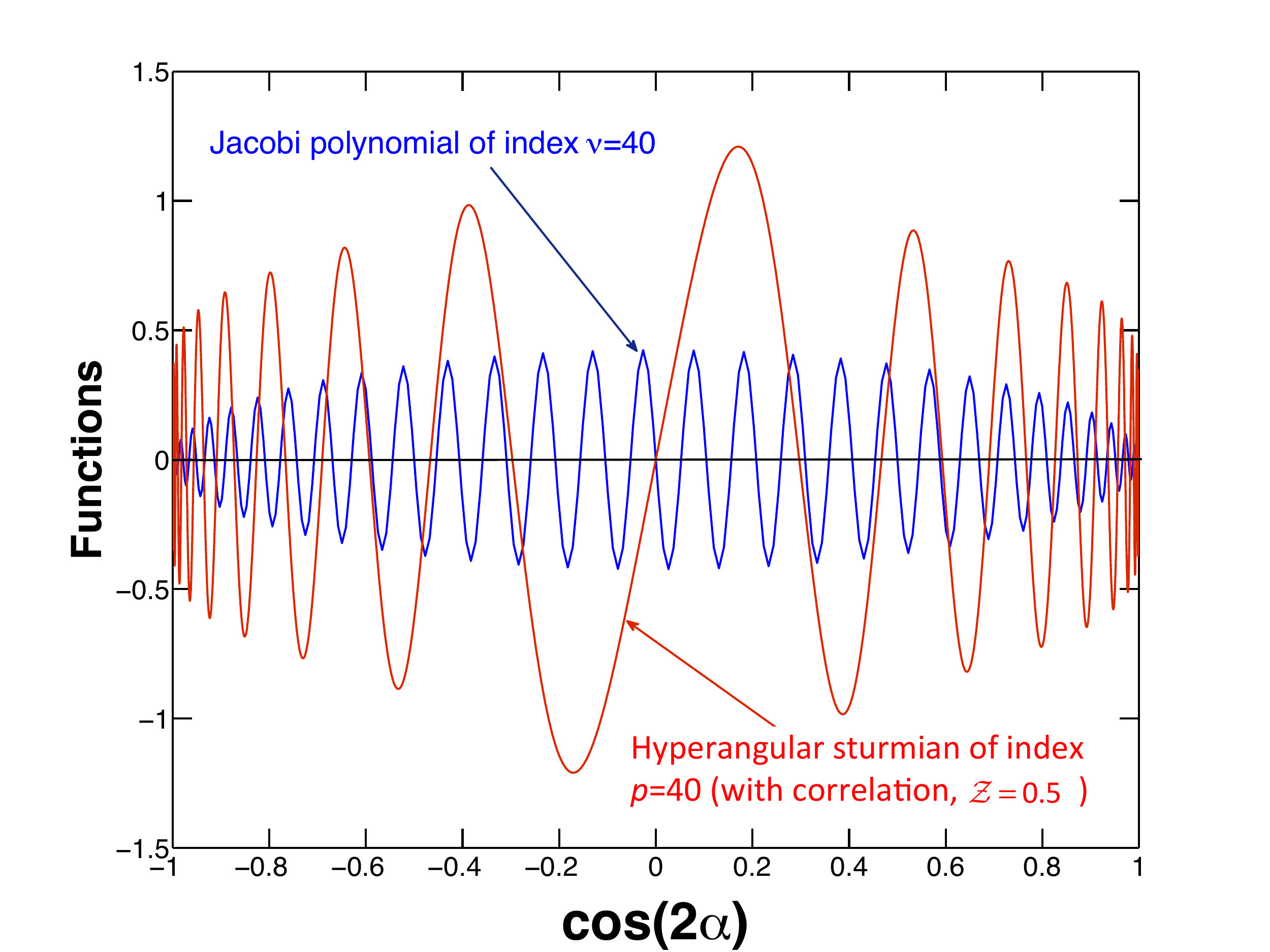}
\caption{ (Color on line) Plot of the Jacobi polynomial of index 40 (blue line) and  the hyperangular sturmian function of index 40 and $\mathcal{Z}=0.5$ for $l_1=l_2=0$ as a function of $\cos(2\alpha)$. 503 B-splines have been used to generate the hyperangular sturmian function. }
\label{Fig2}
\end{figure}
\begin{figure}[h]
\includegraphics[width=12cm,height=9cm]{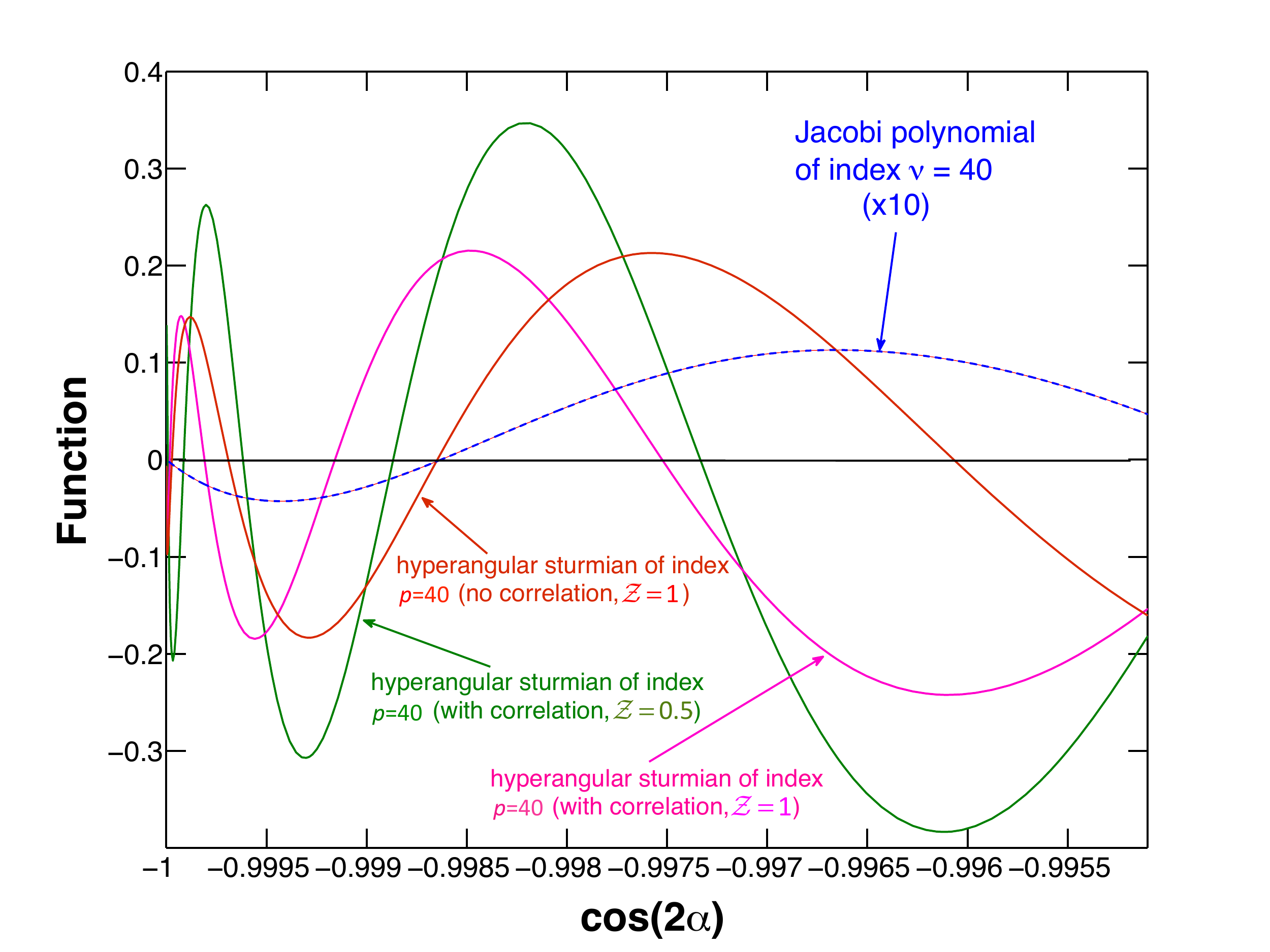}
\caption{(Color on line) Plot of the Jacobi polynomial (multiplied by 10) of the same index 40 and $l_1=l_2=0$ as a function of $\cos{2\alpha}$ very close to $\cos(2\alpha)=-1$ (blue dashed line). This Jacobi polynomial is compared to three hyperangular sturmian functions of index 40 and $l_1=l_2=0$: the red line corresponds to $\mathcal{Z}=1$ within the independent model; the green line corresponds to $\mathcal{Z}=0.5$ with correlation; the magenta line corresponds to $\mathcal{Z}=1$ with correlation. As in Table II, 503 B-splines have been used to generate the hyperangular sturmians.}
\label{Fig3}
\end{figure}

In order to illustrate the above discussion, Fig.(\ref{Fig2}) shows the Jacobi polynomial of index 40 and compares it to the hyperangular sturmian of index 40 and $\mathcal{Z}=0.5$ for $l_1=l_2=0$. We clearly see
that in the case of the Jacobi polynomial, the nodes are distributed quasi uniformly along the $\alpha$ axis between $\alpha=0$ and $\alpha=\pi/2$. This contrasts strongly with  the hyperangular sturmian function the nodes of which are clearly concentrated around $\alpha=0$ and $\alpha=\pi/2$. In Fig.(\ref{Fig3}), we show the behavior of the same Jacobi polynomial very close to $\alpha=\pi/2$ {\it i.e.} $\cos(2\alpha)=-1$ and compare it to the behavior of the same hyperangular sturmian as before and two other hyperangular sturmians, one corresponding to $\mathcal{Z}=1$ and the other one obtained within the independent electron model. As expected, the density of nodes close to $\pi/2$ increases when $\mathcal{Z}$ decreases.\\

In the above discussion, we mentioned the fact that our hyperangular sturmian functions take into account at least partially the electron-electron correlation. For that reason, we expect our hyperangular sturmian function basis to be appropriate to describe the electron-electron dynamics. This, is the basic brick of the three-body problem. This brings our basis closer to the fully correlated basis as those defined in \cite{Rodriguez2005,Rodriguez2007,Gasaneo2008,Pekeris,Wintgen,Binwei,Grosges} but using hyperspherical coordinates. However, we must stress that our approach does not fulfill the Kato cusp condition associated with the two-electron coalescence. This means that the eigenenergy of mainly the ground state is limited to 5 or 6 digits. Note that for our purpose, this is sufficient. The other states are much less affected by this Kato cusp condition since the electrons are most of the time far from each other.\\

\subsubsection{Hyperradial sturmian functions}
Let us now consider the hyperradial sturmian functions $S_{n,p}(R)$. In the system of the hyperspherical coordinates, the Schr\"odinger  equation (\ref{Reduced}) has the same structure as the equation for  a hydrogen-like system. It is possible to use Generalized sturmian functions as done in \cite{Gasaneo2013}, however, as suggested in \cite{Gasaneo2012} for hyperspherical coordinates it is  convenient to use Coulomb sturmian functions that are solution of the following Sturm-Liouville problem:
\begin{equation}
\left[-\frac{1}{2}\left(\frac{\mathrm{d}^2}{\mathrm{d}R^2}-\frac{\ell(\ell+1)}{R^2}\right)-\frac{\kappa v}{R}+\frac{\kappa^2}{2}\right]S_{v,\ell}^{\kappa}(R)=0,  \label{RadialStur}
\end{equation}
with the following boundary conditions:
\begin{eqnarray}
\lim_{R\rightarrow 0}S_{v,\ell}^{\kappa}(R)&=& 0,\\
\lim_{R\rightarrow\infty}S_{v,\ell}^{\kappa}(R)&=& 0.
\end{eqnarray}
In equation (29), $\ell$ and $v$ are now half integers \cite{Rem} given by:
\begin{eqnarray}
\ell&=&\lambda'+3/2, \\ 
v&=&n+3/2, 
\end{eqnarray}
where $\lambda'$ and $n$ are positive integers with $n\geq \lambda'+1$. $\kappa v$ is the eigenvalue where $\kappa$ is an arbitrarily fixed parameter which acts like a spatial dilatation factor. The choice of the value of $\kappa$ will be discussed later. Eq. (\ref{RadialStur}) may then be rewritten as follows:
\begin{equation}
\left[-\frac{1}{2}\left(\frac{\mathrm{d}^2}{\mathrm{d}R^2}-\frac{\lambda'(\lambda'+4)+\frac{15}{4}}{R^2}\right)-\frac{\kappa(n+\frac{3}{2})}{R}+\frac{\kappa^2}{2}\right]S_{n,\lambda'}^{\kappa}(R)=0.  \label{RadialStur1}
\end{equation}
Note that the centrifugal terms in Eqs. (\ref{AngSturmian}) and (\ref{RadialStur}) are similar. $\lambda$ and $\lambda'$ can then be chosen identical to remove the centrifugal term from Eq. (14). The analytical expression of the Coulomb sturmian function is:
\begin{equation}
S_{n,\lambda'}^{\kappa}(R)= N_{n,\lambda'}^{\kappa}R^{\lambda'+\frac{5}{2}}e^{-\kappa R} L_{n-\lambda'-1}^{2\lambda'+4}(2\kappa R), 
\end{equation}
where $L_{n-\lambda'-1}^{2\lambda'+4}(2\kappa R)$ is a Laguerre polynomial. The normalization factor $N_{n,\lambda'}^{\kappa}$ given by:
\begin{equation}
N_{n,\lambda'}^{\kappa}=\sqrt{\frac{\kappa}{n+3/2}}\;(2\kappa)^{\lambda'+5/2}\left[\frac{(n-\lambda'-1)!}{(n+\lambda'+3)!}\right]^{\frac{1}{2}},
\end{equation}
is derived from the condition:
\begin{equation}
\int_0^{\infty}S_{n,\lambda'}^{\kappa}(R)S_{n,\lambda'}^{\kappa}(R)\;\mathrm{d}R=1.
\end{equation}
These Coulomb sturmian functions form a complete and discrete set of functions that are orthogonal with respect to a weight function which is the Coulomb potential $1/R$.\\

Masili and Starace \cite{Starace3}  found that including logarithmic terms in $R$ in their basis expansions involving Slater orbitals  improved convergence for a non optimised basis in calculating energy levels and the static and dynamic polarizabilities  of helium.  They also found that using an optimised basis of Slater orbitals in $R$ without logarithmic terms gave similar accuracy.  We do not include logarithmic terms in $R$ in our basis  but use an optimised basis by varying the $\kappa$ parameter and using more than one $\kappa$. 

\subsubsection{Oscillator strengths}
In order to assess the accuracy of both the energy and the wave function of the bound states obtained with our approach, it is instructive to calculate the corresponding oscillator strengths. They are expressed  in terms of the electric-dipole matrix elements. Here, we assume that the electric field is linearly pola\-rized along the the z-axis. For a transition from an initial state $|\Phi_i^{L,M}\rangle$ to a final state $|\Phi_f^{L',M'}\rangle$, these oscillator strengths are defined in the length $(L)$ and the velocity $(V)$ gauge as follows:\\
\begin{eqnarray}
f_L&=&C(E_f-E_i)\left|\langle\Phi_f^{L',M'}|\vec{e}_z\cdot(\vec{r}_1+\vec{r}_2)|\Phi_i^{L,M}\rangle\right|^2,\\  \nonumber \\
f_V&=&\frac{C}{(E_f-E_i)}\left|\langle\Phi_f^{L',M'}|\vec{e}_z\cdot(\vec{\nabla}_1+\vec{\nabla}_2)|\Phi_i^{L,M}\rangle\right|^2.\\ \nonumber
\end{eqnarray}
In expressions (38) and (39), $E_i$ and $E_f$ are the energies of the initial and final states  and $C$ is a constant equal to 2 and 5/3 for $S\rightarrow P$ and $P\rightarrow D$ transitions respectively. In practice, we first calculate the length and velocity gauge dipole matrix in our sturmian basis and then obtain the oscillator strength by moving from the sturmian basis to the atomic basis. The detailed calculation of the dipole matrix elements in the sturmian basis is lengthy but straightforward. We briefly describe this calculation in the appendix.

\subsection{Numerical implementation of the sturmian basis}

\subsubsection{The hyperangular sturmian basis}

To construct the hyperangular sturmian functions $H^{l_1,l_2}_{p,\mathcal{Z},\lambda}(\alpha)$ in equation (\ref{AngSturmian}) we expand them in terms of B-splines, $B_i(\alpha)$,\cite{Bachau}
\begin{equation}
H_{p,\mathcal{Z},\lambda}^{{l_1}{l_2}}\left( \alpha  \right) = \sum\limits_{i = 1}^n {c_i^p} {B_i}\left( \alpha  \right).\label{Bspline}
\end{equation}
It is straightforward to enforce the boundary conditions in equations $(23)$ and  (\ref{BounConditions}) by having all of the $B_i$ equal to zero at $\alpha=0$ and $\alpha=\pi/2.$ The B-splines are a very flexible basis set allowing any form of mesh one would like in $\alpha$ and hence allowing for an accurate description of the $H^{l_1,l_2}_{p,\mathcal{Z},\lambda}(\alpha)$ near $\alpha=0$ and $\alpha=\pi/2$ where there are an increasing number of oscillations. Fixing the values of $\lambda, l_1$ and $l_2$ in eq. (\ref{AngSturmian}) and substituting the expansion in (\ref{Bspline}) leads to a generalized eigenvalue problem for the eigenvalues $\rho_p(\lambda)$ and the eigenvector components $c_i^p$. \\

\begin{figure}[h!]
\includegraphics[width=12cm,height=9cm]{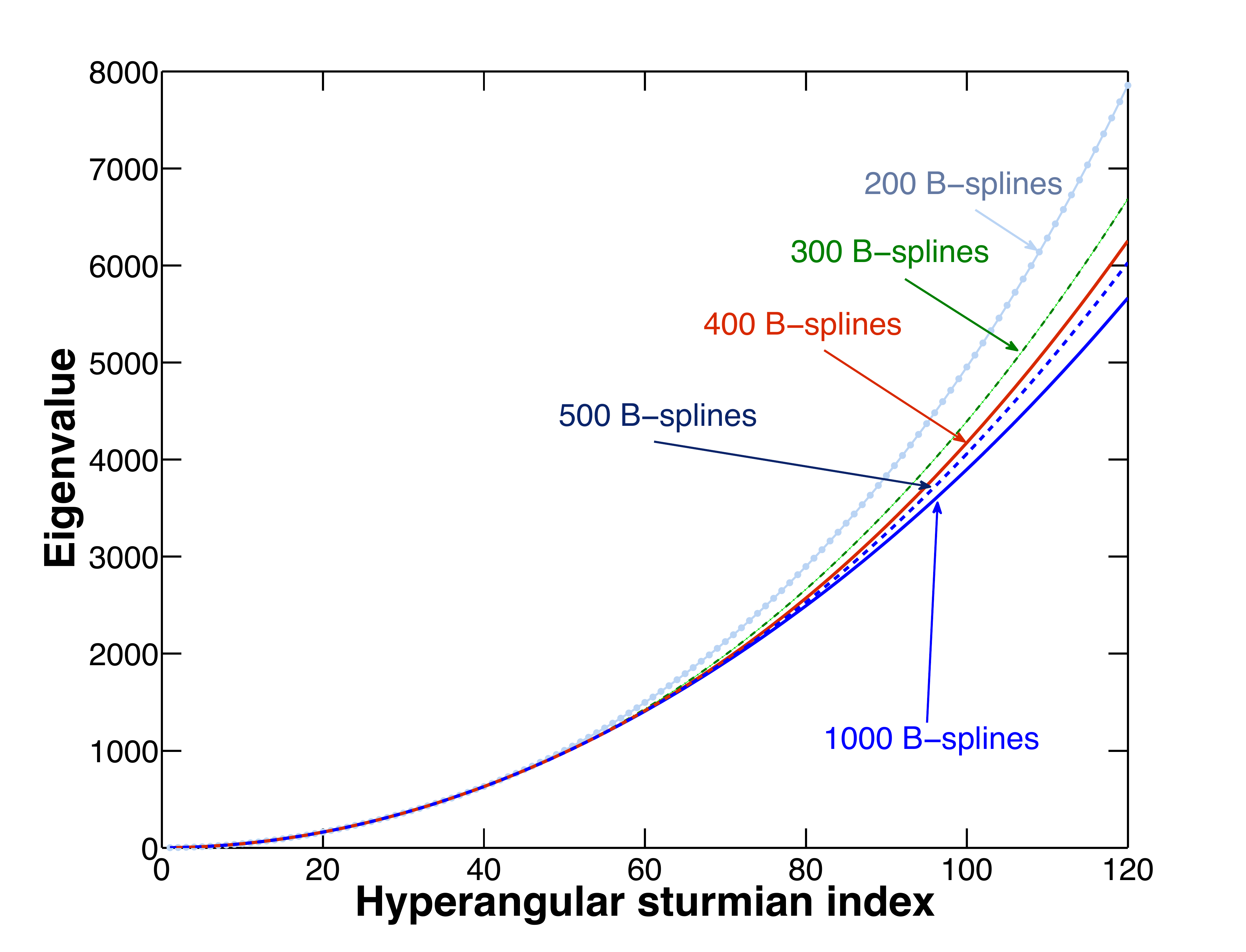}
\caption{(Color on line) Eigenvalue of the corresponding hyperangular sturmian $H^{0,0}_p(\alpha)$ eigenfunction solution of the Sturm-Liouville problem (22). The various curves correspond to different 
                  sizes of the B-spline basis in term of which these hyperangular sturmian functions are expanded.}
\label{Fig4}
\end{figure}

For $l_1 = l_2$ the solutions of eq. (\ref{AngSturmian}) are either even or odd with respect to reflection about $\alpha=\pi/4$ and so in the sum in eq. (\ref{Expansion}) only the even solutions in $\alpha$ contribute to singlet ($S=0$) states  and the odd solutions for the triplet states ($S=1$). When $l_1$ and $l_2$ are different, for the exchange term in (\ref{Expansion}), we can either use $H^{l_1,l_2}_{p,\mathcal{Z},\lambda}(\frac{\pi}{2}-\alpha)$ or $H^{l_2,l_1}_{p,\mathcal{Z},\lambda}(\alpha)$. Fig. (\ref{Fig4}) gives the eigenvalue associated to each hyperangular sturmian $H^{0,0}_{p,\mathcal{Z},\lambda}(\alpha)$ as a function of the index $p$ for various sizes of the B-spline basis. In all these bases, equally spaced mesh points are used. We see that for the first 50 eigenvalues, the results are stable showing that a basis of 200 B-splines is enough. Note that for our purpose, 50 hyperangular sturmians per 
$(l_1,l_2)$ pair is sufficient in most of the cases. If however higher eigenvalues are needed, Fig. (\ref{Fig4}) shows that bigger sizes of the B-spline basis will be necessary while using a density of mesh points higher near $\alpha=0$ and $\alpha=\pi/2$.\\ \\

\subsubsection{Matrix formulation of the Schr\"odinger equation}

Let us now consider  the matrix formulation of the Schr\"odinger equation (\ref{Reduced}). The solution (\ref{Expansion}) is first expanded in terms of  hyperrangular and hyperradial sturmian functions. After
substitution in equation (\ref{Reduced}) and making use of Eqs. (\ref{AngSturmians1}) and (\ref{AngSturmian}), we obtain:
\begin{eqnarray}
\sum_{n,p,l_1,l_2}\psi^{L,M}_{n,p,l_1,l_2}& &\frac{ \phi_{p,l_1,l_2}^{L,M}(\alpha,\hat r_1,\hat r_2)}{\sqrt{2(1+\delta_{l_1,l_2})}}\;\;
[ - \left( \frac{1}{2}\frac{\partial^2}{\partial R^2}-\frac{\lambda'(\lambda'+4)+\frac{15}{4}}{2R^2}\right)\nonumber\\ 
&-&\frac{2  \rho_p \, \tilde C(\alpha,\hat{r}_1,\hat{r}_2)}{2R^2}+\frac{ C(\alpha,\hat{r}_1,\hat{r}_2)}{R}-E_{\beta}\;\;]\;\;S_{n,\lambda'}^{\kappa}(R)
 =0. \label{Solution2}
\end{eqnarray}
where only one of the terms needed to make symmetric the wave function was included to make the steps more easy to follow.
Using  Eq. (\ref{RadialStur1}) with $\lambda'=\lambda$ gives:
\begin{eqnarray}
\sum_{n,p,l_1,l_2}\psi^{L,M}_{n,p,l_1,l_2}& & \frac{\phi_{p,l_1,l_2}^{L,M}(\alpha,\hat r_1,\hat r_2)S_{n,\lambda}^{\kappa}(R)}{\sqrt{2(1+\delta_{l_1,l_2})}}\;\;
[-\frac{ \rho_p \, \tilde C(\alpha,\hat{r}_1,\hat{r}_2)}{R^2}\nonumber\\
&+&\frac{ C(\alpha,\hat{r}_1,\hat{r}_2)+\kappa(n+\frac{3}{2})}{R} -\left( E_{\beta}+\frac{\kappa^2}{2}\right)\;\;] 
=0, \label{Solution2}
\end{eqnarray}
The hyperangular sturmian equation (\ref{AngSturmian}) provides a set of eigenvalues that can be assimilated to a discretization of the hyperradial coordinates. For $R=\rho_p$, the first term in the above equation cancels exactly the parts of $ C(\alpha,\hat{r}_1,\hat{r}_2)$ included in $ \tilde{C}(\alpha,\hat{r}_1,\hat{r}_2)$. Projecting on the left with the basis functions and integrating we obtain:
\begin{eqnarray}
\sum_{n,p,l_1,l_2}\frac{\psi^{L,M}_{n,p,l_1,l_2}}{\sqrt{2(1+\delta_{l_1,l_2})}}\;\;[
&-&\left[  \frac{ \rho_p }{R^2}\right]_{n',p',n,p} \delta_{\nu',\nu}
+\delta_{n',n} \left[ C(\alpha,\hat{r}_1,\hat{r}_2)+\kappa(n+\frac{3}{2})\right]_{\nu',\nu}\nonumber\\ 
&-&\left[  E_{\beta}+\frac{\kappa^2}{2}\right]_{\tau',\tau}\;\;]  
=0, \label{Solution3}
\end{eqnarray}
where we used $\tau=\left\lbrace n,p,l_1,l_2\right\rbrace $. As the result of  the orthonormality relation of the sturmian functions we have:
\begin{equation}
 \left[\tilde C(\alpha,\hat{r}_1,\hat{r}_2) \right]_{p',l'_1,l'_2,p,l_1,l_2}=\delta_{p',p}\delta_{l'_1,l_1}\delta_{l'_2,l_2}=\delta_{\nu',\nu}, \label{delta}
\end{equation}
\begin{equation}
\left[  \frac{ 1 }{R}\right]_{n',n}=\frac{\kappa}{n+\frac{3}{2}}\delta_{n',n}, \label{delta2}
\end{equation}
It is interesting to note that using the hyperspherical coordinates, the hyperradial and the hyperangular integrations separate making the calculations easier.

\subsubsection{Calculation of the energy and width of the doubly excited states}

The energy and width of the doubly excited states are obtained by diagonalizing the complex scaled atomic hamiltonian. In the case of the hyperspherical coordinates system, it is only the hyperradius $R$ which is affected by the complex scaling
\begin{equation}
R\rightarrow Re^{i\theta},
\end{equation}
where the scaling angle $\theta$ is real and positive. The atomic hamiltonian is no longer hermitian but complex symmetric with complex eigenvalues. The imaginary part of the energy represents half of the width of the corresponding state. For well described bound states, the imaginary part of the energy is zero while it is nonzero and negative for doubly excited states. It is interesting to note that instead of complex scaling the hamiltonian, it is equivalent to complex scaling the nonlinear parameter $\kappa$ of the hyperradial sturmian:
\begin{equation}
\kappa\rightarrow \kappa e^{-i\theta}.
\end{equation}
In this condition, the hyperradial sturmian functions become complex and behave asymptotically as outgoing spherical waves.

\subsubsection{Optimization of the sturmian basis}

Our long term objective is to study the interaction of a two-active electron system with an external field by solving the Time-Dependent Schr\"odinger Equation (TDSE). Within the framework of a spectral method, it is crucial to reduce as much as possible the size of the basis while keeping a good level of accuracy in describing the two-electron wave packet. Before describing an efficient way of optimizing this basis, it is convenient at this stage to compare the present approach to Lagmago's method \cite{Lagmago} which uses a system of spherical coordinates. This latter method  consists in expanding the solution of the TDSE in a basis of products of Coulomb-sturmian functions of the electron radial coordinates and bipolar harmonics of the angular coordinates. This method was extremely efficient to generate accurate values for the energy and width of a wide range of singlet and triplet resonance states of helium \cite{Eiglsperger1,Eiglsperger2}. \\

Lagmago's method has however two important drawbacks. First, the accurate calculation of the matrix associated to the electron-electron Coulomb interaction term which involves a double integration, requires a computer time that becomes prohibitive when many configurations are included. This is in fact the bottleneck of Lagmago's method. By contrast, this calculation in our new approach is trivial. Secondly in Lagmago's method, the density of single continuum states below the double ionization threshold obtained by diagonalizing the atomic hamiltonian is very low even for very large bases. We have checked that this is no longer the case in our new approach. Nevertheless, It has to be stressed that with a relatively small size basis, Lagmago's method allows one to generate very accurate energies for singly and doubly excited states \cite{Eiglsperger2}. This is particularly true when the level of excitation of both electrons is strongly asymmetric. In this latter case, the electronic cloud is characterized by two distinct regions of space where each electron state is practically coulombic, one close to the nucleus for the inner electron and a region at large distances for the outer electron. An efficient expansion for the wave function should then contain both distance scales and span the two regions simultaneously \cite{Lagmago}. This is achieved by associating to each electron a different value of the nonlinear parameter $\kappa$  (see Eq. (35)) in each product of Coulomb-sturmian functions. For a given atomic state, good values of the nonlinear parameters $\kappa$ are consistently obtained by exploiting the fact that the Coulomb-sturmian  function $S^{\kappa_i}_{n,\ell}(r_i)$ $(i=1, 2$ for electron 1 and 2 respectively)  describes exactly an electron of energy $\epsilon=-\kappa_i^2/2=-z^2/n^2$ in the field of a nucleus of charge $z$ \cite{Foumouo}. In order to describe accurately many atomic states with a single basis, it is therefore necessary to introduce various pairs $(\kappa_1,\kappa_2)$ of nonlinear parameters. This, however, makes the basis numerically overcomplete. It means that some of the eigenvalues of the overlap matrix are very close or equal to 0 because the corresponding eigenvectors are linearly dependent. In \cite{Foumouo}, it is explained in detail how to eliminate these eigenvalues leading to a significant reduction -typically about 30 percents- of the size of the basis while giving a very good description of the electronic structure of the atom.\\

The sturmian basis that we developed  in the system of hyperspherical coordinates can also be optimized by means of a very similar method. The hyperradial and hyperangular sturmian functions depend on the two nonlinear parameters $\kappa$ and $\mathcal{Z}$ respectively. The idea is therefore to introduce several pairs $(\kappa,\mathcal{Z})$ of nonlinear parameters within a single basis. This raises the question of the choice of these pairs of nonlinear parameters. In expression (35) of the hyperradial sturmian function, $\kappa$ appears as a factor that scales the hyperradius $R$ which in turns defines the size of the atom. A large $\kappa$ is therefore used to describe accurately very compact atomic states as for instance the ground state while small values of $\kappa$ will be more convenient for the description of excited states of the atom. In practice, we proceed as follows. For the compact ground state of helium, we take $\kappa\in [1,2]$. Varying the value of $\kappa$ in this interval hardly changes the value of the ground state energy for a fixed number of hyperradial sturmians. If we need to describe accurately asymmetrically excited states, it is the level of excitation of the most excited electron which determines the value of $\kappa$. Following what is done in \cite{Lagmago}, we choose $\kappa=1/n$ where $n$ is the principal quantum number of the most excited electron, assuming both electrons independent. When many excited states of the atom have to be well described simultaneously, we introduce several values of $\kappa$ within the same basis. The other nonlinear parameter $\mathcal{Z}$ is associated to the hyperangle $\alpha$ which in turn controls the relative distance of the electrons with respect to the nucleus. $\mathcal{Z}$ can be interpreted as a weighted mean of the nuclear charges experienced by both electrons. If  electron 1 is close to the nucleus while electron 2 is far from it, both electrons experience different nuclear charges: $Z$ for electron 1 and $Z-\sigma$ for electron 2 where $\sigma$ results from screening by the inner electron. In practice, we choose $\mathcal{Z}=2$  for the ground state of helium. As a matter of fact, the ground state energy of helium is not very sensitive to the value of $\mathcal{Z}$. However, for asymmetrically excited states of helium, it is important to choose a value of $\mathcal{Z}$ close to 0.5 which is the smallest possible value in order to take into account the fact that one electron may be very close to the nucleus while the other one is far, thereby requiring a good description of the wave function around $\alpha=0$ or $\pi/2$. It is precisely for $\mathcal{Z}=0.5$ that the number of nodes of the hyperangular sturmian is the highest around $\alpha=0$ and $\pi/2$. 

\section{Results and discussions}

In the absence of electron-electron interaction, the system of hyperspherical coordinates "introduces" artificially radial electronic correlations. It is therefore interesting in this case, to compare the convergence of the energy of various bound states of helium without the electron-electron interaction term as a function of the number $N$ of Jacobi polynomials or hyperangular sturmians  used in the two corresponding bases. The results are given in Table 1 for the ground state energy and the energy of the $1s2s$ and $2p^2$ states. In the case of our sturmian basis,  
\begin{table}[h]
\caption{Convergence of the ground state energy as well as the energy of the first and second excited states of He within the independent electron model as a function of the number N of  hyperangular sturmian functions and, for comparison, of the number of Jacobi polynomials included in the basis. The exact value of each energy is given at the bottom of each column. All energies are expressed in a.u.}
\vspace{0.5cm}
\begin{tabular}{p{0.4cm}p{2.2cm}p{2.3cm}p{2.2cm}p{2.3cm}p{2.2cm}p{2.3cm}}\hline \hline \\
 &\multicolumn{2}{c} {$-E_{1s^2}\;\;\;\;\;\;\;\;$} & \multicolumn{2}{c} {$-E_{1s2s}\;\;\;\;\;\;\;\;$}&\multicolumn{2}{c} {$-E_{2p^2}\;\;\;\;\;\;\;\;$}\\ \\
\hline \\
N&Hyperangular sturmians&Jacobi polynomials&Hyperangular sturmians&Jacobi polynomials&Hyperangular sturmians&Jacobi polynomials \\  \\ \hline \\
5&3.9999999231&3.9900995024&2.4997988721&2.4055497334&0.9999999636&0.9994140601\\
6&3.9999999990&3.9939504282&2.4999886272&2.4332192268&0.9999999991&0.9997154676\\
7&4.0000000000&3.9960485770&2.4999994522&2.4512870831&0.9999999999&0.9998496994\\
10&4.0000000000&3.9985612968&2.5000000000&2.4782054468&0.9999999999&0.9999682508\\
20&4.0000000000&3.9998096196&2.5000000000&2.4964050795&0.9999999999&0.9999987677\\
30&4.0000000000&3.9999427995&2.5000000000&2.4988565368&0.9999999999&0.9999998565\\
40&4.0000000000&3.9999757880&2.5000000000&2.4995031283&0.9999999999&0.9999999922\\
50&4.0000000000&3.9999876339&2.5000000000&2.4997415786&0.9999999999&1.0000000217\\
80&4.0000000000&3.9999970915&2.5000000000&2.4999356353&0.9999999999&1.0000000350\\
100&4.0000000000&3.9999985912&2.5000000000&2.4999668851&0.9999999999&1.0000000360\\
110&4.0000000000&3.9999989832&2.5000000000&2.4999750907&0.9999999999&1.0000000362\\ \\
&\multicolumn{2}{c} {\textbf{4.0000000000}} & \multicolumn{2}{c} {\textbf{2.5000000000}}&\multicolumn{2}{c} {\textbf{1.0000000000}}\\ \\
\hline \hline
\end{tabular}
\label{tab.1}
\end{table}
we used 10 hyperradial sturmians and set $\kappa=2$ and $\mathcal{Z}=2$. For the ground state energy, we see that the convergence is reached with only 7 hyperangular sturmians while more than 100 Jacobi polynomials are not sufficient to reach convergence. The same conclusions hold in the case of the $1s2s$ and $2p^2$ states. A close look at the $1s2s$ state indicates that in the case of the Jacobi polynomials, the convergence of the energy toward the correct value is extremely slow thereby demonstrating the real efficiency of our new sturmian approach.\\

\begin{table}[h]
\caption{ Convergence of the absolute value of the ground state energy in a.u. of H$^-$ as a function of the number of ($l_1,l_2$) pairs included in the basis and of the basis size.  The nonlinear parameters $\kappa=0.8$ and $\mathcal{Z}=0.7$. The number of hyperangular and hyperradial sturmians is 5  for the results in the third column and 8 and 7 respectively for the results given in the fifth column.  The data in the last column are from \cite{Foumouo} and the accurate value of this energy has been taken from \cite{Drake}.\\}.
\vspace{0.5cm}
\begin{tabular}{p{1cm}p{1cm}p{3cm}p{1cm}p{3cm}p{1cm}p{3cm}}\hline\hline \\
&\multicolumn{6}{c} {Absolute value of the ground state energy of H$^-$} \\
$(l_1,l_2)$&size &present results&size&present results&size&reference \cite{Foumouo}\\  \\ \hline \\
$(0,0)$&25&0.5143565816&56&0.5144929336&465&0.51449614\\
$(1,1)$&50&0.5264960358&112&0.5265826368&930&0.52658410\\
$(2,2)$&75&0.5273498205&168&0.5274362924&1395&0.52743744\\
$(3,3)$&100&0.5275368039&224&0.5276233311&1860&0.52762391\\
$(4,4)$&125&0.5275997809&280&0.5276863687&2325&0.52768618\\
$(5,5)$&150&0.5276266394&336&0.5277132796&2790&0.52771215\\
$(6,6)$&175&0.5276399520&392&0.5277266356 \\       
$(7,7)$&200&0.5276472766&448&0.5277339957 \\
$(8,8)$&225&0.5276516285&504&0.5277383767  \\ \\
&\multicolumn{6}{c} {Accurate value \cite{Drake}: \textbf{0.5277510165443}} \\
\hline\hline
\end{tabular}
\label{tab.2}
\end{table}

In Table 2, we analyze the convergence of the ground state energy of H$^-$, including the electron-electron interaction term, as a function of the number of $(l_1,l_2)$ pairs of electron angular momenta. We present two sets of results. For the first set, the number of hyperangular and hyperradial  sturmians is equal to 5 while it is equal to 8 and 7 respectively for the second one. In both cases, $\kappa =0.8$ and $\mathcal{Z}=0.7$. Our results are compared to those of Foumouo {\it et al.} who used a system of spherical coordinates  and a basis of products of Coulomb sturmians of the radial coordinates and bipolar harmonics of the angular coordinates. For completeness, we also give a very accurate value of this energy obtained by Drake \cite{Drake}. The first point to underline is the very small number of hyperangular and hyperradial sturmians needed to get a relatively accurate result for the ground state energy of H$^-$. By contrast, the sturmian method based on the spherical coordinates \cite{Foumouo} requires bases of much bigger size. The second point to underline is the slow convergence of this ground state energy as a function of the number of $(l_1,l_2)$ pairs. It is important to stress that the present approach like all approaches  of configuration-interaction type do not satisfy the Kato cusp condition associated with the coalescence of the two electrons \cite{Kato}. This leads to a slow convergence which is only acute however in the case of the ground state where both electrons are strongly confined. By contrast, correlated bases in which the basis functions depend explicitly on the interelectronic distances, do satisfy the Kato cusp condition. These bases which usually require prohibitively large matrix sizes, give very accurate results.  For our final purpose namely the time-propagation of a two-electron wave packet, the accuracy reached in the present calculations of the ground state energy is more than enough.\\

\begin{table}[h]
\caption{ Convergence of the absolute value of the ground state energy in a.u. of He as a function of the number of $(l_1,l_2)$ pairs included in the basis and of the basis size.  The nonlinear parameters are $\kappa=2$ and $\mathcal{Z}=2$. The number of hyperangular and hyperradial sturmians is 5. The present results are compared to the data from \cite{Gerard} and \cite{Foumouo}. The accurate value of the ground state energy has been taken from \cite{Burgers}.}
\vspace{0.5cm}
\begin{tabular}{p{1cm}p{1cm}p{3cm}p{1cm}p{3cm}p{1cm}p{3cm}}\hline\hline \\
&\multicolumn{6}{c} {Absolute value of the ground state energy of He} \\ \\
$(l_1,l_2)$&size &present results&size&reference \cite{Gerard}&size&reference \cite{Foumouo}\\ \\  \hline \\
$(0,0)$&25&2.8790259232&55&2.8790102261&465&2.87902797\\
$(1,1)$&50&2.9005099765&110&2.9004681981&930&2.90051386\\
$(2,2)$&75&2.9027595331&165&2.9026813178&1395&2.90276209\\
$(3,3)$&100&2.9033130359&220&2.9031912216&1860&2.90331321\\
$(4,4)$&125&2.9035099926&275&2.9033435028&2325&2.90350682\\
$(5,5)$&150&2.9035966612&330&2.9033917089&2790&2.90358925\\
$(6,6)$&175&2.9036404813&385&2.9034066146&3255&2.90362816\\
$(7,7)$&200&2.9036649390&440&2.9034109991&6560&2.90366100\\
$(8,8)$&225&2.9036796027&495&2.9034122278&&\\ \\
&\multicolumn{6}{c} {Accurate value \cite{Burgers}: \textbf{2.9037243770341}} \\ \\
\hline\hline
\end{tabular}
\label{tab.3}
\end{table}
\begin{table}[h!]
\caption{ Absolute value of the energy in a.u. of the first five singly excited singlet states of helium for L=0, 1 and 2. (a), the present results, (b) results from \cite{Gerard} and (c) results from \cite{Burgers} for L=0 and from \cite{Drake2} for L=1 and 2. The present results are obtained with 10 hyperradial sturmians with $\kappa=1$ and 40 with $\kappa=0.3$  and 6 hyperangular sturmians with $\mathcal{Z}=0.7$ and 44  with $\mathcal{Z}=0.5$. Five pairs $(l_1,l_2)$ of electron angular momenta are included.}
\vspace{0.5cm}
\begin{tabular}{p{1cm}p{2.5cm}p{2.5cm}p{2.5cm}p{2.5cm}p{2.5cm}}\hline\hline \\
&\multicolumn{1}{c}{2S}&\multicolumn{1}{c}{3S}&\multicolumn{1}{c}{4S}&\multicolumn{1}{c}{5S}&\multicolumn{1}{c}{6S}\\ \\
(a)&2.14596163060&2.06126853687&2.03358530496&2.02117613863&2.01456263146\\
(b)&2.14594146031&2.06125528815&2.03357048378&2.02115898556&2.01453730633\\
(c)&2.1459740461&2.0612719897&2.0335867169&2.0211768512&2.0145630974\\  \\ \hline \\
&\multicolumn{1}{c}{2P}&\multicolumn{1}{c}{3P}&\multicolumn{1}{c}{4P}&\multicolumn{1}{c}{5P}&\multicolumn{1}{c}{6P}\\ \\
(a)&2.12383270946&2.05514301634&2.03106820074&2.01990523641&2.01383351214\\
(b)&2.12382134535&2.05513631281&2.03106096815&2.01989933948&2.01382673587\\
(c)&2.1238430865&2.0551463621&2.0310696505&2.0199059899&2.0138339797\\  \\ \hline \\
&\multicolumn{1}{c}{3D}&\multicolumn{1}{c}{4D}&\multicolumn{1}{c}{5D}&\multicolumn{1}{c}{6D}&\multicolumn{1}{c}{7D}\\ \\
(a)&2.05562014201&2.03127951254&2.02001564409&2.01389808784&2.01020972397\\
(b)&2.05561876719&2.03127873641&2.02001501876&2.01389746634&2.01020932503\\
(c)&2.0556207329&2.0312798462&2.0200158362&2.0138982274&2.0102100285\\ \\
\hline \hline
\end{tabular}
\label{tab.4}
\end{table}

\begin{table}[h]
\caption{ Absolute value of the energy in a.u. of the first five singly excited triplet states of helium for L=0, 1 and 2. (a), the present results, (b) results from \cite{Gerard} and (c) results from \cite{Burgers}. The same sturmian basis as in Table 4 has been used. }
\vspace{0.5cm}
\begin{tabular}{p{1cm}p{2.5cm}p{2.5cm}p{2.5cm}p{2.5cm}p{2.5cm}}\hline\hline \\
&\multicolumn{1}{c}{2S}&\multicolumn{1}{c}{3S}&\multicolumn{1}{c}{4S}&\multicolumn{1}{c}{5S}&\multicolumn{1}{c}{6S}\\ \\
(a)&2.17522862118&2.06868889674&2.03651201827&2.02261883997&2.01537741998\\
(b) &2.175227505&2.068688594&2.036511897&2.022618781&2.015377402\\
(c)&2.17522937824&2.06868906747&2.03651208310&2.02261887230&2.01537745299\\  \\ \hline \\
&\multicolumn{1}{c}{2P}&\multicolumn{1}{c}{3P}&\multicolumn{1}{c}{4P}&\multicolumn{1}{c}{5P}&\multicolumn{1}{c}{6P}\\ \\
(a) &2.13316338940&2.05808086502&2.03232426495&2.02055114068&2.01420790853\\
(b)&2.1331604001&2.0580771526&2.0323220232&2.0205495485&2.0142057491\\
(c)&2.13316419078&2.05808108427&2.03232435430&2.02055118726&2.01420795877\\  \\ \hline \\
&\multicolumn{1}{c}{3D}&\multicolumn{1}{c}{4D}&\multicolumn{1}{c}{5D}&\multicolumn{1}{c}{6D}&\multicolumn{1}{c}{7D}\\ \\
(a) &2.05562444264&2.03128358627&2.02001826151&2.01389976915&2.01021084673\\
(b)&2.0556360463&2.0312886903&2.0200208815&2.0139012356&2.0102118550\\
(c)&2.05563630945&2.03128884750&2.02002102745&2.01390141545&2.01021210596\\ \\
\hline \hline
\end{tabular}
\label{tab.5}
\end{table}

\begin{table}[h]
\caption{ Absolute value of the energy in a.u. and width in a.u. of the lowest doubly excited states of H$^-$ for L=0, 1 and 2. The present results are shown
in the last two columns. They are obtained with 20 hyperradial sturmians with $\kappa=0.8$, 20 with $\kappa=0.5$ and 20 with $\kappa=0.1$ as
well as 10 hyperangular sturmians with $\mathcal{Z}=0.7$ and 40 with $\mathcal{Z}=0.5$. Five $(l_1,l_2)$ pairs are included in the calculations. Our 
present results are compared to data, the reference of which is indicated in the Table, and which have been obtained, for all of them, by means of an 
explicitly correlated basis. }
\vspace{0.5cm}
\begin{tabular}{p{3cm}p{2cm}p{2cm}p{2cm}p{2cm}p{2.2cm}}\hline\hline\\
\multicolumn{2}{c} {Reference data} & \multicolumn{2}{c} {Reference \cite{Scrinzi}} & \multicolumn{2}{c} {Present results} \\ \\
Energy&Width&Energy&Width&Energy&Width\\  \\ \hline \\ 
 \multicolumn{6}{c} {\textbf{L=0}} \\ \\
\cite{Ho1} 0.1487765 &1.731(-3)&0.1487762&1.7332(-3)&0.1487759&1.73398(-3)\\
\cite{Ho2} 0.1260196 &8.985(-5)&0.1260199&9.02(-5)&0.1260200&9.057(-5)\\
&&0.069006&1.4184(-3)&0.0690059&1.4189(-3)\\
&&0.0561434&8.8(-5)&0.0561407&8.907(-5)\\
&&&&&\\ 
\multicolumn{6}{c} {\textbf{L=1}} \\ \\
\cite{Ho2} 0.1260495&1.165(-6)&0.12604986&1.36(-6)&0.1260498&1.3645(-6)\\
\cite{Ho3} 0.12436&6.9(-4)&0.1243856&7.0(-4)&0.1243645&6.573(-4)\\
\cite{Ho4} 0.06871675&1.1914(-3)&0.062716&1.19(-3) &0.0627167&1.1909(-3)\\
\cite{Ho4} 0.0585718&8.986(-6)&0.0585718&8.988(-6)&0.0585718&8.9873(-6)\\ \\
\multicolumn{6}{c} {\textbf{L=2}} \\ \\
\cite{Ho2} 0.12794175&3.1625(-4)&0.127937&3.12(-4)&0.1279156&3.1769(-4)\\
\cite{Bhatia} 0.0659533&1.6581(-3)&0.0659531&1.6576(-3)&0.0659382&1.6612(-3)\\
&&0.0568294&2.5302(-4)&0.0568251&2.5376(-4)\\ \\ \hline \hline
\end{tabular}
\label{tab.6}
\end{table}
\begin{table}[h]
\caption{ Oscillator Strength corresponding to $S\rightarrow P$ transitions in helium. Our results which are obtained within the length gauge
                  are compared to accurate data given in the literature. }
\vspace{0.5cm}
\begin{tabular}{p{4.2cm}p{3cm}p{3cm}p{3cm}p{2.8cm}}\hline\hline\\

&HSCC-method\;\cite{Tang}&variational method\;\cite{Hijikata}&Variational method\;\cite{Schiff}&Present method\\ 
Transition&length gauge&length gauge&length gauge&length gauge\\ \\ \hline\\
$1s1s\; ^1S^e\rightarrow 1s2p\; ^1P^0$&0.2762&0.2761 &0.2761&0.276089\\
$1s1s\; ^1S^e\rightarrow 1s3p\; ^1P^0$&0.07429&0.0736 &0.074&0.073417\\
$1s1s\; ^1S^e\rightarrow 1s4p\; ^1P^0$&0.03022&0.0301&0.030&0.029941\\
$1s2s\; ^1S^e\rightarrow 1s2p\; ^1P^0$&0.3774&0.3760&0.3764&0.376483\\
$1s2s\; ^1S^e\rightarrow 1s3p\; ^1P^0$&0.1494&0.1486&0.1514&0.151325\\
$1s2s\; ^1S^e\rightarrow 1s4p\; ^1P^0$&0.0515&0.0521&0.049&0.049286\\
$1s3s\; ^1S^e\rightarrow 1s2p\; ^1P^0$&-0.1437&-0.1379&-0.1453&-0.145489\\
$1s3s\; ^1S^e\rightarrow 1s3p\; ^1P^0$&0.6206&0.6221&0.626&0.626260\\
$1s3s\; ^1S^e\rightarrow 1s4p\; ^1P^0$&0.1435&0.1465&0.144&0.144260\\
\\
\hline
\end{tabular}
\label{tab.7}
\end{table}


In Table 3, we consider the ground state energy of He and study its convergence as a function of the number of $(l_1,l_2)$ pairs. The nonlinear parameters are $\kappa=2$ and $\mathcal{Z}=2$. The number of hyperangular and hyperradial sturmians is equal to 5. As in the previous case, the size of the basis needed to get a relatively accurate result is significantly smaller than in the case where the spherical coordinate system is used \cite{Gerard,Foumouo}. The same conclusions as above regarding the slow convergence of the results as a function of the number of pairs $(l_1,l_2)$ hold.\\

In Table 4, we present results for the energies of the first five singly excited singlet states of He for L=0, 1 and 2. These data result from a single diagonalization of the hamiltonian matrix in our sturmian basis. In the present case, it is necessary to optimize the basis by introducing several set $(\kappa,\mathcal{Z})$ of nonlinear parameters. We use 10 hyperradial sturmians with $\kappa=1$ and 40 with $\kappa=0.3$ as well as 6 hyperangular sturmians with $\mathcal{Z}=0.7$ and 44 with $\mathcal{Z}=0.5$. We have included five $(l_1,l_2)$ pairs of electron angular momenta. Our results are compared to Lagmago's results \cite{Gerard} who used an optimized sturmian basis in the spherical coordinate system. Note that this latter method is particularly suitable to calculate the energy of high asymmetrically excited states with an accuracy  similar to the one obtained with correlated bases. In Table 4 where we consider low-lying excited states, we see that the present results are in most of the cases slightly more accurate than Lagmago's results in the sense that they are closer to the value obtained with correlated bases (see \cite{Burgers} for S states and \cite{Drake2} for P and D states). The results for singly excited triplet states are presented in Table 5 where we use the same sturmian basis as in Table 4. Note that  Pauli's principle prevents the coalescence of the two electrons to occur. As a result, the fact that the Kato cusp condition is not fulfilled has a weaker impact on the accuracy of the results. From Table 5, we see that the results obtained with the present approach are very close to Lagmago's results \cite{Gerard} and compare rather well with the accurate data given in \cite{Burgers}.\\ 

In Table 6, we consider the first doubly excited states of H$^-$ for $L=0, 1$ and $2$. The width of these states is defined as twice the imaginary part of the corresponding eigenenergy of the complex scaled hamiltonian. As for the singly excited states of He, several pairs of nonlinear parameters have to be used within the same basis to generate many accurate energies within a single diagonalization. However, it is important at this stage to pay attention to the following point. By contrast to atoms, the spectrum of negative ions contains shape resonances which do not reduce to doubly excited bound states when the electron-electron interaction term is artificially switched off. In that case, the choice of adequate values for $\kappa$ is no longer obvious. We checked that small values of this nonlinear parameters have to be included in the basis. In addition, since the width of  these shape resonances is rather large, the angle $\theta$ of complex scaling must be relatively large. In the present case, $\theta=0.25$ in radians and we used 20 hyperradial sturmians  with $\kappa=0.8$, 20 with $\kappa=0.5$ and 20 with $\kappa=0.1$ as well as 10 hyperangular sturmians with $\mathcal{Z}=0.7$ and 40 with $\mathcal{Z}=0.5$. For the results presented in Table 6, we checked that it is sufficient to take five $(l_1,l_2)$ pairs into account. Our results are compared to data which have been obtained, for all of them, by means of explicitly correlated bases. We clearly see on Table 6 that the present basis which is not explicitly correlated in the sense that it does not satisfy the Kato cusp condition, provides rather accurate results, even for shape resonances, that compare very well with those obtained with explicitly correlated bases.\\

Finally, we present in Table 7 our results for oscillator strengths corresponding to $S\rightarrow P$ transitions in helium. These calculations, which have been performed in the length gauge, allow us to assess the accuracy of the bound state wave functions at large distances and provide a stringent test of the quality of the wave functions. In our calculations, we take five $(l_1,l_2)$ pairs into account. We use within the same basis, two values of the hyperradial sturmian nonlinear parameter $\kappa$ ($\kappa_1=1$ and $\kappa_2=0.1$) with 20 hyperradial sturmians for each value of $\kappa$. We use 30 hyperangular sturmians with $\mathcal{Z}=0.6$. Despite the fact that the total size of our basis is rather small, our results compare very well with  data obtained with accurate variational methods and explicitly correlated bases \cite{Hijikata,Schiff}. We have also a good agreement with the results of Tang {\it et al.} who used the so-called Hyperspherical Close-Coupling (HSCC) method \cite{Tang2} which is based on a efficient treatment of the adiabatic expansion introduced by Macek \cite{Macek1}.

\section{Conclusion and perspective}
In this contribution, we have developed an efficient  {\it ab initio} spectral approach to calculate the energy spectrum of a two-active electron atom in a system of hyperspherical coordinates. The key point of this approach is the introduction  in our basis of new  sturmian functions of the hyperangle. These functions which form a complete set, are solution of a Sturm-Liouville equation that treats exactly the electron-nucleus interaction and the radial electron-electron correlation term. As a result, the size of the  hyperangular sturmian basis needed for an accurate description of most of the energy eigenfunctions is remarkably small. For the hyperradial part of our basis, it is equally convenient to use Coulomb sturmians since in terms of the hyperradius, the problem is purely coulombic. In addition, both the hyperangular and the hyperradial sturmians depend on two nonlinear parameters namely a weighted mean effective charge $\mathcal{Z}$ and a wave vector $\kappa$ respectively. By introducing various sets $(\mathcal{Z},\kappa)$ of nonlinear parameters which are chosen on physical grounds we are able to generate a large number of atomic state energies through a single diagonalization of the hamiltonian while significantly reducing the size of the basis. In order to assess the efficiency of the present method  we  have calculated the energy of the ground state and various excited states of He as well as  the ground state energy and the energy and width of the first doubly excited states of H$^-$ for three values of the total angular momentum. Our results compare very well with the data provided by other approaches including  those which use an explicitly correlated basis. The same conclusions apply to the calculation of oscillator strengths in helium. By using the length gauge, we showed that the accuracy of the bound state wave functions generated with the present method is very good even at large distances.\\

Our main objective is to use this spectral approach to solve the time-dependent Schr\"odinger equation to treat the interaction of atoms with strong laser fields. Within this context, this approach has two major advantages. First, it can be generalized easily to the treatment of atomic systems with more than two active electrons. Indeed, the atomic hamiltonian keeps exactly the same structure as for atomic hydrogen in which the electrostatic potential is replaced by an effective charge function of various hyperangles divided by the hyperradius. Second, irrespective of the number of active electrons, there is always only one coordinate, namely the hyperradius, which is unbound. This makes  easier the implementation of various methods such as t-SURFF \cite{Scrinzi2} and the time-scaled coordinate method \cite{Hamido,Frapiccini} aimed at extracting from the final many electron wave packet the information on the electron energy spectra. However by contrast to the electronic structure calculations, the Coulomb sturmian functions are less adapted to describe accurately the continua. In this case, B-splines, finite elements or even finite difference techniques may be easily implemented.

\section{Acknowledgements}
A.A. and B.P. thank Laurence Malegat for enlightening discussions about the problems related to the expansion in terms of hyperspherical harmonics.
G.G. thanks the Universit\'e Catholique de Louvain (UCL) for financially supporting several stays at the Institute of Condensed Matter and Nanosciences of the UCL. F.M.F and P.F.O'M gratefully acknowledge the European network COST (Cooperation in Science and Technology) through the Action CM1204 "XUV/X-ray light and fast ions for ultrafast chemistry" (XLIC) for financing several short term scientific missions at UCL. 

\section{Appendix}
In this appendix, we give the general expression of the electric-dipole matrix elements in our sturmian basis. More details about the calculation of some of the expressions presented here are given in \cite{Starace2}. The electric-dipole matrix elements are denoted by $D_L$ in the length gauge and by $D_V$ in the velocity gauge. Their 
integral  expression is:\\
\begin{eqnarray}
D_L=\int\frac{2}{R^{5/2}\sin 2\alpha}& &S_{n',\lambda}^{\kappa}(R)H_{p',\mathcal{Z},\lambda}^{l_1',l_2'}(\alpha)Y_{l_1',l_2'}^{L',M'}(\Omega_1,\Omega_2)\;\left[\vec{e}_z\cdot(\vec{r}_1+\vec{r}_2)\right]\nonumber\\
&&\frac{2}{R^{5/2}\sin 2\alpha}S_{n,\lambda}^{\kappa}(R)H_{p,\mathcal{Z},\lambda}^{l_1,l_2}Y_{l_1,l_2}^{L,M}(\Omega_1,\Omega_2)\;dV,
\end{eqnarray}
\begin{eqnarray}
D_V=\int\frac{2}{R^{5/2}\sin 2\alpha}& &S_{n',\lambda}^{\kappa}(R)H_{p',\mathcal{Z},\lambda}^{l_1',l_2'}(\alpha)Y_{l_1',l_2'}^{L',M'}(\Omega_1,\Omega_2)\;\left[\vec{e}_z\cdot(\vec{\nabla}_1+\vec{\nabla}_2)\right]\nonumber\\
&&\frac{2}{R^{5/2}\sin 2\alpha}S_{n,\lambda}^{\kappa}(R)H_{p,\mathcal{Z},\lambda}^{l_1,l_2}Y_{l_1,l_2}^{L,M}(\Omega_1,\Omega_2)\;dV,\\ \nonumber
\end{eqnarray}
where the volume element $dV$ is given by:
\begin{equation}
dV=R^5\sin^2\alpha\cos^2\alpha\;dR\;d\alpha\;d\Omega_1\;\Omega_2.\\ \nonumber
\end{equation}
In order to simplify the notations, it is convenient to define the following function:
\begin{equation}
\Psi_{n,p,l_1,l_2}^{L,M}(R,\alpha,\Omega_1,\Omega_2)=\frac{2}{R^{5/2}\sin 2\alpha}S_{n,\lambda}^{\kappa}(R)H_{p,\mathcal{Z},\lambda}^{l_1,l_2}Y_{l_1,l_2}^{L,M}(\Omega_1,\Omega_2).\\
\end{equation}
Let us start with the length gauge. We first define the two following terms:
\begin{equation}
H_{L,M,L',M',p,p'}^{l_1,l_2,l_1',l_2'}=\langle H_{p',\mathcal{Z},\lambda}^{l_1',l_2'}|\cos\alpha|H_{p,\mathcal{Z},\lambda}^{l_1,l_2}\rangle\langle Y_{l_1',l_2'}^{L',M'}|\cos\theta_1|Y_{l_1,l_2}^{L,M}\rangle,
\end{equation}
\begin{equation}
K_{L,M,L',M',p,p'}^{l_1,l_2,l_1',l_2'}=\langle H_{p',\mathcal{Z},\lambda}^{l_1',l_2'}|\sin\alpha|H_{p,\mathcal{Z},\lambda}^{l_1,l_2}\rangle\langle Y_{l_1',l_2'}^{L',M'}|\cos\theta_2|Y_{l_1,l_2}^{L,M}\rangle.\\
\end{equation}
In this case, the electric-dipole matrix element $D_L$ is:
\begin{eqnarray}
D_L&=&\langle \Psi_{n',p',l_1',l_2'}^{L',M'} |R(\cos\alpha\cos\theta_1+\sin\alpha\cos\theta_2)|\Psi_{n,p,l_1,l_2}^{L,M}\rangle,\nonumber\\
&=&\langle S_{n',\lambda}^{\kappa}|R|S_{n,\lambda}^{\kappa}\rangle\left(H_{L,M,L',M',p,p'}^{l_1,l_2,l_1',l_2'}+K_{L,M,L',M',p,p'}^{l_1,l_2,l_1',l_2'}\right).
\end{eqnarray}
where we use the usual Dirac notations.
The angular factors present in expressions (52) and (53) are easy to calculate. They can be expressed in terms of the following terms:\\
\begin{equation}
A_{L,M,L',M'}^{l_1,l_2,l_2'}=\delta_{l_2',l_2}(-1)^{l_2-M'}\sqrt{(2L+1)(2L'+1)(2l_1+1)}\left(\begin{array}{ccc}
                                                                                                                               L  &  L'  &  1 \\
                                                                                                                              -M &  M' &  0
                                                                                                                          \end{array}\right),
\end{equation}
\begin{equation}
B_{L,M,L',M'}^{l_1,l_1',l_2}=\delta_{l_1',l_1}(-1)^{l_1-M'}\sqrt{(2L+1)(2L'+1)(2l_2+1)}\left(\begin{array}{ccc}
                                                                                                                               L  &  1 &  L' \\
                                                                                                                              -M &  0 &  M'
                                                                                                                          \end{array}\right),
\end{equation}
\begin{equation}
C_{L,L'}^{l_1,l_1',l_2}(l)=\delta_{l_1',l}\sqrt{(2l+1)}\left(\begin{array}{ccc}
                                                                                                      l & 1 & l_1 \\
                                                                                                     0 & 0 & 0
                                                                                                 \end{array}\right)
                                                                                         \left\{\begin{array}{ccc}
                                                                                                     L & L' & 1 \\
                                                                                                     l & l_1 & l_2
                                                                                                 \end{array}\right\},
\end{equation}
\begin{equation}
E_{L,L'}^{l_1,l_2,l_2'}(l)=\delta_{l_2',l}\sqrt{(2l+1)}\left(\begin{array}{ccc}
                                                                                                      l & 1 & l_2 \\
                                                                                                     0 & 0 & 0
                                                                                                 \end{array}\right)
                                                                                         \left\{\begin{array}{ccc}
                                                                                                     L & 1 & L' \\
                                                                                                     l & l_1 & l_2
                                                                                                 \end{array}\right\}. \\ \nonumber
\end{equation}
We have:\\
\begin{eqnarray}
\langle Y_{l_1',l_2'}^{L',M'}|\cos\theta_1|Y_{l_1,l_2}^{L,M}\rangle&=&A_{L,M,L',M'}^{l_1,l_2,l_2'}\left(C_{L,L'}^{l_1,l_1',l_2}(l_1+1)+C_{L,L'}^{l_1,l_1',l_2}(l_1-1)\right),\\
\langle Y_{l_1',l_2'}^{L',M'}|\cos\theta_2|Y_{l_1,l_2}^{L,M}\rangle&=&B_{L,M,L',M'}^{l_1,l_1',l_2}\left(E_{L,L'}^{l_1,l_2,l_2'}(l_2+1)+E_{L,L'}^{l_1,l_2,l_2'}(l_2-1)\right).\\ \nonumber
\end{eqnarray}
Let us now consider the velocity gauge:\\
\begin{eqnarray}
D_V&=&\langle \Psi_{n',p',l_1',l_2'}^{L,M}|(\cos\alpha\cos\theta_1+\sin\alpha\cos\theta_2)\frac{\partial}{\partial R}|\Psi_{n,p,l_1,l_2}^{L,M}\rangle\nonumber\\
&&+\langle \Psi_{n',p',l_1',l_2'}^{L,M}|\sin^2\alpha\cos\theta_1\frac{1}{R}\;\;\frac{\partial}{\partial \cos\alpha}|\Psi_{n,p,l_1,l_2}^{L,M}\rangle\nonumber\\
&&-\langle \Psi_{n',p',l_1',l_2'}^{L,M}|\cos\alpha\sin\alpha\cos\theta_2\frac{1}{R}\;\;\frac{\partial}{\partial \cos\alpha}|\Psi_{n,p,l_1,l_2}^{L,M}\rangle\nonumber\\
&&+\langle \Psi_{n',p',l_1',l_2'}^{L,M}|\frac{\sin^2\theta_1}{R\cos\alpha}\;\;\frac{\partial}{\partial\cos\theta_1}|\Psi_{n,p,l_1,l_2}^{L,M}\rangle\nonumber\\
&&+\langle \Psi_{n',p',l_1',l_2'}^{L,M}|\frac{\sin^2\theta_2}{R\sin\alpha}\;\;\frac{\partial}{\partial\cos\theta_2}|\Psi_{n,p,l_1,l_2}^{L,M}\rangle.\\ \nonumber
\end{eqnarray}
For the sake of clarity, let us define the following factors:\\
\begin{equation}
R_{n,n'}=\langle S_{n',\lambda}^{\kappa}|\frac{\partial}{\partial R}|S_{n,\lambda}^{\kappa}\rangle-\frac{5}{2}\langle S_{n',\lambda}^{\kappa}|\frac{1}{R}|S_{n,\lambda}^{\kappa}\rangle,
\end{equation}
\begin{equation}
S_{n,n'}=\langle S_{n',\lambda}^{\kappa}|\frac{1}{R}|S_{n,\lambda}^{\kappa}\rangle,
\end{equation}
\begin{equation}
L_{L,M,L',M',p,p'}^{l_1,l_2,l_1',l_2'}=\langle H_{p',\mathcal{Z},\lambda}^{l_1',l_2'}|\frac{\cos 2\alpha}{\cos\alpha}|H_{p,\mathcal{Z},\lambda}^{l_1,l_2}\rangle\langle Y_{l_1',l_2'}^{L',M'}|\cos\theta_1|Y_{l_1,l_2}^{L,M}\rangle,
\end{equation}
\begin{equation}
P_{L,M,L',M',p,p'}^{l_1,l_2,l_1',l_2'}=\langle H_{p',\mathcal{Z},\lambda}^{l_1',l_2'}|\sin\alpha\frac{\partial}{\partial\alpha}|H_{p,\mathcal{Z},\lambda}^{l_1,l_2}\rangle\langle Y_{l_1',l_2'}^{L',M'}|\cos\theta_1|Y_{l_1,l_2}^{L,M}\rangle,
\end{equation}
\begin{equation}
Q_{L,M,L',M',p,p'}^{l_1,l_2,l_1',l_2'}=\langle H_{p',\mathcal{Z},\lambda}^{l_1',l_2'}|\frac{\cos 2\alpha}{\sin\alpha}|H_{p,\mathcal{Z},\lambda}^{l_1,l_2}\rangle\langle Y_{l_1',l_2'}^{L',M'}|\cos\theta_2|Y_{l_1,l_2}^{L,M}\rangle,
\end{equation}
\begin{equation}
T_{L,M,L',M',p,p'}^{l_1,l_2,l_1',l_2'}=\langle H_{p',\mathcal{Z},\lambda}^{l_1',l_2'}|\cos\alpha\frac{\partial}{\partial\alpha}|H_{p,\mathcal{Z},\lambda}^{l_1,l_2}\rangle\langle Y_{l_1',l_2'}^{L',M'}|\cos\theta_2|Y_{l_1,l_2}^{L,M}\rangle,
\end{equation}
\begin{equation}
U_{L,M,L',M',p,p'}^{l_1,l_2,l_1',l_2'}=\langle H_{p',\mathcal{Z},\lambda}^{l_1',l_2'}|\frac{1}{\cos\alpha}|H_{p,\mathcal{Z},\lambda}^{l_1,l_2}\rangle\langle Y_{l_1',l_2'}^{L',M'}|\sin\theta_1\frac{\partial}{\partial\theta_1}|Y_{l_1,l_2}^{L,M}\rangle,
\end{equation}
\begin{equation}
V_{L,M,L',M',p,p'}^{l_1,l_2,l_1',l_2'}=\langle H_{p',\mathcal{Z},\lambda}^{l_1',l_2'}|\frac{1}{\sin\alpha}|H_{p,\mathcal{Z},\lambda}^{l_1,l_2}\rangle\langle Y_{l_1',l_2'}^{L',M'}|\sin\theta_2\frac{\partial}{\partial\theta_2}|Y_{l_1,l_2}^{L,M}\rangle.\\  
\end{equation}
The new angular factors which appear in the above expressions write:\\
\begin{eqnarray}
\langle Y_{l_1',l_2'}^{L',M'}|\sin\theta_1\frac{\partial}{\partial\theta_1}|Y_{l_1,l_2}^{L,M}\rangle&=&l_1A_{L,M,L',M'}^{l_1,l_2,l_2'}C_{L,L'}^{l_1,l_1',l_2}(l_1+1)\nonumber\\
&&-(l_1+1)A_{L,M,L',M'}^{l_1,l_2,l_2'}C_{L,L'}^{l_1,l_1',l_2}(l_1-1),
\end{eqnarray}
\begin{eqnarray}
\langle Y_{l_1',l_2'}^{L',M'}|\sin\theta_2\frac{\partial}{\partial\theta_2}|Y_{l_1,l_2}^{L,M}\rangle&=&l_2B_{L,M,L',M'}^{l_1,l_1',l_2}E_{L,L'}^{l_1,l_2,l_2'}(l_2+1)\nonumber\\
&&-(l_2+1)B_{L,M,L',M'}^{l_1,l_1',l_2}E_{L,L'}^{l_1,l_2,l_2'}(l_2-1).
\end{eqnarray}
The final expression of $D_V$ becomes:\\
\begin{eqnarray}
D_V&=&R_{n,n'}\left(H_{L,M,L',M',p,p'}^{l_1,l_2,l_1',l_2'}+K_{L,M,L',M',p,p'}^{l_1,l_2,l_1',l_2'}\right)\nonumber\\
&&+S_{n,n'}\left[\left(L_{L,M,L',M',p,p'}^{l_1,l_2,l_1',l_2'}-P_{L,M,L',M',p,p'}^{l_1,l_2,l_1',l_2'}\right)-\left(Q_{L,M,L',M',p,p'}^{l_1,l_2,l_1',l_2'}-T_{L,M,L',M',p,p'}^{l_1,l_2,l_1',l_2'}\right)\right]\nonumber\\
&&-S_{n,n'}\left(U_{L,M,L',M',p,p'}^{l_1,l_2,l_1',l_2'}+V_{L,M,L',M',p,p'}^{l_1,l_2,l_1',l_2'}\right).\\ \nonumber
\end{eqnarray}
Note that the matrix elements involving the hyperradial sturmians may be calculated analytically. However, when various values of the nonlinear $\kappa$ parameter are
used, these matrix elements have to be calculated numerically by means of a Gauss-Laguerre quadrature which provides an exact results if the number of points is sufficient.

\smallskip

\begin{thebibliography}{9}
\bibitem{Rem} This contrasts with what we have in the system of spherical coordinates where $v$ and $\ell$ are pure integers.
\end{thebibliography}
\end{document}